\documentclass[lettersize,journal,10pt]{IEEEtran}
\usepackage{amsmath,amssymb,amsfonts}
\usepackage{algorithmic}
\usepackage{algorithm}
\usepackage{array}
\usepackage{textcomp}
\usepackage{stfloats}
\usepackage{url}
\usepackage{verbatim}
\usepackage{graphicx}
\usepackage{cite}
\usepackage{xcolor}
\usepackage{dcolumn}
\usepackage{makecell}
\usepackage{multirow}
\usepackage{enumerate}
\usepackage{subfloat}
\usepackage{float}
\usepackage{booktabs}
\usepackage[caption=false,font=footnotesize,labelfont=rm,textfont=rm]{subfig}
\allowdisplaybreaks[4]

\makeatletter
\renewcommand{\maketag@@@}[1]{\hbox{\m@th\normalsize\normalfont#1}}%
\makeatother

\makeatletter
\renewcommand{\citepunct}{,\penalty\@m\hskip.13emplus.1emminus.1em}
\renewcommand{\citedash}{\hbox{--}\penalty\@m}

\begin{document}
\begin{sloppypar}

\title{Learning of Uplink Resource Allocation with Multiuser QoS Constraints}

\author{
	\IEEEauthorblockN{Manru Yin, Shengqian Han, and Chenyang Yang}
	\\ \IEEEauthorblockA{Beihang University, Beijing, China\\
		Email: \{mryin, sqhan, cyyang\}@buaa.edu.cn}
\vspace{-0.8cm}
}

\maketitle

\begin{abstract}
In the paper the joint optimization of uplink multiuser power and resource block (RB) allocation are studied, where each user has quality of service (QoS) constraints on both long- and short-blocklength transmissions. The objective is to minimize the consumption of RBs for meeting the QoS requirements, leading to a mixed-integer nonlinear programming (MINLP) problem. We resort to deep learning to solve the problem with low inference complexity. To provide a performance benchmark for learning based methods, we propose a hierarchical algorithm to find the global optimal solution in the single-user scenario, which is then extended to the multiuser scenario. The design of the learning method, however, is challenging due to the discrete policy to be learned, which results in either vanishing or exploding gradient during neural network training. We introduce two types of smoothing functions to approximate the involved discretizing processes and propose a smoothing parameter adaption method. Another critical challenge lies in guaranteeing the QoS constraints. To address it, we design a nonlinear function to intensify the penalties for minor constraint violations. Simulation results demonstrate the advantages of the proposed method in reducing the number of occupied RBs and satisfying QoS constraints reliably.
\end{abstract}

\section{Introduction}
The fifth-generation (5G) and beyond systems are designed to support diverse quality of service (QoS) requirements~\cite{Song2024QoS,bairagi2020coexistence}. For uplink transmission, jointly optimizing the allocation of users' transmit power and the system's resource blocks (RBs) is crucial to guarantee the QoS requirements for both long-blocklength transmission (LBT) and short-blocklength transmission (SBT). 
This optimization problem, however, falls into the framework of mixed integer and nonlinear programming (MINLP)  \cite{Chen2021Joint,zarini2023resource,alsenwi2022coexistence}.
Its nondeterministic polynomial hard (NP-hard) nature makes it difficult to find an optimal solution. Traditional tools often require significant computational effort to solve such problems, which are difficult to meet the real-time requirement of practical applications. Deep learning has been adopted to learn the relationship between environmental parameters and the  optimized resource allocation. With a well-trained neural network, learning based methods offer the advantage of low inference complexity in decision making~\cite{sun2022learning,Guo2023Deep}.

A primary challenge in applying learning based methods for MINLP  problems is the discrete nature of the policy to be learned. 
Neural network parameters are typically updated via gradient descent algorithms, while the discretizing process that converts the continuous output of deep neural networks (DNNs) to discrete variables incurs either vanishing or exploding gradients.
Four kinds of methods have been proposed to address this issue in the literature. The first method employs the straight-through estimator (STE), which simply sets the gradient of non-differentiable functions used for  discretizing process as one during gradient backpropagation~\cite{van2017neural}. However, while efficient in computation, this method entails gradient mismatch, affecting training performance. The second method relaxes discrete variables into continuous counterparts and then introduces additional constraints to ensure that these variables take only discrete values~\cite{zarini2023resource,Hanna2020Distributed}.
This method shifts the challenge from learning discrete policy to ensuring that the learned policy satisfies  constraints. As we will see later, making the learned policy satisfy constraint is not a simple task.
The third method approximates discretizing process with smoothing functions like the Sigmoid function \cite{Fang2020Stochastic} or projection networks \cite{feng2023uplink}.
This method inevitably introduces approximation errors to the original functions, which may degrade the learning performance.
Existing methods introduce smoothing parameters to control approximation errors. Typically, these parameters either remain constant throughout training or are adjusted using an annealing approach~\cite{Fang2020Stochastic,jang2016categorical,maddison2016concrete}.
For instance, the Gumbel-Softmax reparameterization method employs a smoothing parameter, called temperature, within the SoftMax function, which gradually decreases across training iterations based on an annealing method~\cite{jang2016categorical,maddison2016concrete}.
Yet, tuning these smoothing parameters in complex problems is a time-consuming task. 
The fourth method applies the  reinforcement learning, such as deep Q-networks (DQN). DQN estimates the value of each discrete action combination, and subsequently selects the optimal action based on its value~\cite{mnih2015human}. However, as the dimensionality of actions increases, the number of action combinations grows exponentially, complicating the learning process. To tackle this obstacle, a branching dueling Q-network (BDQ) was introduced in \cite{tavakoli2018action}, which achieves a linear growth in the number of action combinations.

Another key challenge lies in ensuring the learned resource allocation policy to satisfy the QoS constraints. Designing an activation function for the output layer of a neural network that satisfies the constraints is a straightforward approach~\cite{liang2019towards}. However, its applicability is largely limited to simple constraints, like ensuring a non-negative power by leveraging activation functions with a non-negative value range.
In \cite{li2021multicell}, a projection block is cascaded after a traditional DNN to satisfy complicated rate constraints. Another widely used approach treats constraints as penalties added to the loss function, with weights serving as fine-tuned hyper-parameters~\cite{liang2019towards,lee2022deep}.
Nonetheless, determining the appropriate penalty weights is demanding, as they affect both the performance and constraint satisfaction. An alternative approach uses Lagrangian dual optimization, where Lagrangian multipliers are learned by the dedicated \emph{multiplier DNNs}~ \cite{Eisen2019Learning,Sun2019Learning}. This approach allows for dynamic adjustment of multipliers. However, the constraints cannot be ensured to be satisfied, which is problematic, especially for applications with strict reliability requirements. To alleviate the learning demands on multiplier DNNs, conservative constraints are often considered during training. For instance, increasing reliability requirements during training can reduce the violation of constraints for the learned policy~\cite{Sun2019Learning}.

In this paper we optimize the joint allocation of users' transmit power and the system's RBs for multiuser uplink transmission, where each user has QoS requirements for~both LBT and SBT. The objective is to minimize the RB consumption to satisfy the QoS requirements of users, thereby conserving RB resources to accommodate other potential uplink requests. The formulated resource allocation problem is an MINLP problem. We propose a learning based method to solve this discrete optimization while satisfying the stringent QoS requirements. The main contributions are summarized as follows:
\begin{itemize}
	\item \textit{Hierarchical algorithm for optimal resource allocation:}
	The studied MINLP resource allocation problem suffers from  prohibitively high complexity using exhaustive searching. To solve the problem and provide a performance benchmark for learning based methods, we propose a hierarchical algorithm to find the global optimal solution in the single-user scenario, which is with a drastically reduced computational complexity. We further extend this algorithm to the multiuser scenario. 
	\item \textit{Adaptive smoothing approximation:} When employing deep learning to solve discrete optimization problems, the training of DNNs through gradient backpropagation performs poorly due to vanishing or exploding gradients. To tackle this issue, we develop an adaptive smoothing method to approximate the discretizing process, where the smoothing parameters are dynamically adjusted based on the updated output of DNNs at every iteration. Such a dynamic adjustment helps in controlling approximation errors and gradient magnitudes of the smoothing functions, thereby enhancing the learning performance.
	\item \textit{Nonlinear penalty for constraint violations:} The learning methods based on Lagrangian dual optimization suffer from slow updates in the trainable parameters of multiplier DNNs when constraints are nearly satisfied. This hinders the effective penalization of slightly violated constraints. This motivates us to design a nonlinear function to intensify penalties for minor constraint violations. By incorporating the nonlinear function with the proposed adaptive smoothing method,  the constraints can be satisfied with high reliability.
\end{itemize}

The remainder of this paper is organized as follows\footnote{A part of this work, specifically the numerical algorithm for single-user  scenario, was presented in a conference paper~\cite{YMR2022PIMRC}. The material in this journal version is substantially extended, including the numerical algorithm for multiuser scenarios, the learning based method, and new simulations results for performance and constraint violation comparisons.}.
The system model and problem formulation are described in Section~II. The numerical optimization method and the learning based method are presented in Section~III and Section~IV, respectively.
In Section~V, we evaluate the performance of the proposed algorithms. Finally, the conclusion is drawn in Section~VI.

\section{System Model and Problem Formulation} \label{Sec: system_model}

\subsection{System Model}
\begin{figure}[h!]
	\centering
	\includegraphics[width=0.8\linewidth]{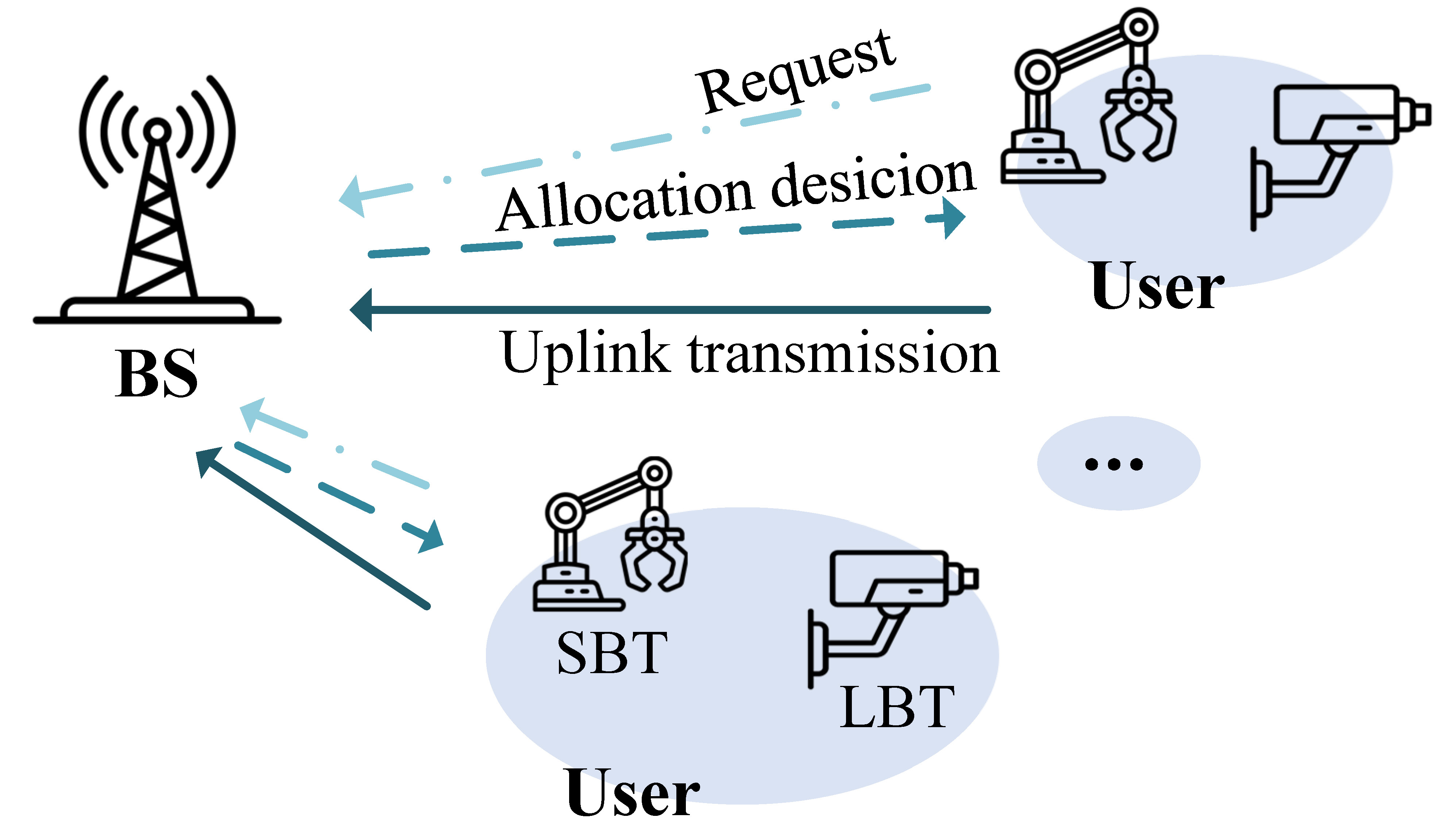}
	\caption{Uplink transmission with the optimization of resource allocation.}
	\label{fig:system_model}
\end{figure}

Consider the uplink transmission scenario of $M$ users, where each user has both LBT and SBT data demands, as shown in Fig.~\ref{fig:system_model}. For example, in a 5G-empowered industry internet of things system, a customer premise equipment (CPE) may concurrently upload surveillance videos and remote control signals~\cite{huaweiWP}.
The considered scenario can be straightforwardly simplified when considering only LBT or~SBT.

The base station (BS) is equipped with $N_r$ receive antennas and each user is equipped with a single antenna. The transmissions of users are scheduled on different RBs of an orthogonal frequency division multiple access (OFDMA) system to eliminate inter-user interference.
Consider that the channel state information of users is available at the BS. The BS optimizes the resource allocation and informs the users of the allocation decisions for transmission.

The achievable rate for LBT can be modeled by the Shannon capacity. In contrast, for SBT, the capacity in the short-blocklength regime \cite{marco2014second} should be considered to model the achievable rate, which is
\begin{equation}
\begin{aligned}
   & r^S_m = LB\sum_{f=1}^F x_{m,f}^S \log ( 1+ \gamma_{m,f} p_{m,f}) \\  &-\frac{1}{\tau} \sqrt{LB\tau\sum\nolimits_{f=1}^F x_{m,f}^S \left(1-\frac{1}{\left(1+ \gamma_{m,f} p_{m,f}\right)^2}\right)}Q^{-1}(\epsilon_m), \label{short_urllc0}
\end{aligned}
\end{equation}
where $r^S_m$ is the achievable rate for SBT of user~$m$, $F$ is the total number of RBs, $L$ is the number of subcarriers in each RB, $B$ is the subcarrier spacing, $\tau$ is the duration of a time slot, 
$x_{m,f}^{S}\in\{0, 1\}$ indicates whether RB$_f$ is assigned to the SBT of user~$m$,  $\gamma_{m,f}=\frac{\alpha}{\sigma^2}\|\mathbf{h}_{m,f}\|^2$ represents the noise-normalized effective channel gain of user~$m$ in RB$_f$ with maximum ratio combining at the BS (called channel gain for short in the sequel),
$\alpha$ is the large-scale channel gain, $\sigma^2$ is the noise power,  $\mathbf{h}_{m,f}\in\mathbb{C}^{N_r\times 1}$ is the channel vector of user~$m$ in RB$_f$, which is assumed to remain constant across the subcarriers in RB$_f$ over the duration $\tau$, $p_{m,f}$ is the power allocated to RB$_f$ by user $m$, and
$Q^{-1}\left(\cdot\right)$ is the inverse of $Q$~function, and $\epsilon_m$ is the decoding error probability of user~$m$. The dispersion $1-\frac{1}{\left(1+ \gamma_{m,f} p_{m,f}\right)^2}$ in \eqref{short_urllc0} is upper bounded by $1$. The bound is tight when the channel gain $\gamma_{m,f}$ is large~\cite{schiessl2015delay}, which is often the common case in 5G and beyond systems equipped with large antenna arrays at the BS.

\subsection{Problem Formulation}
We optimize the joint allocation of users' transmit powers and the system's RBs to meet the QoS requirements for both LBT and SBT. For user $m$, the QoS requirement for LBT is characterized by the minimal rate requirement $R^L_m$, while the QoS requirement for SBT is characterized by both the minimal rate requirement $R^S_m$ and the decoding error probability $\epsilon_m$, as defined in \eqref{short_urllc0}.
The objective is to minimize the RB consumption, thereby conserving RB resources to accommodate other potential uplink requests.

The optimization problem can be formulated~as
\begin{subequations}\label{initial_problem}
	\begin{align}
	 {\!\!\!\!}\min_{ p_{m,f},\atop x_{m,f}^{L},x_{m,f}^{S}}&\!\! \sum_{m=1}^{M}\sum_{f=1}^{F} \left(x_{m,f}^{L} +x_{m,f}^{S} \right) \label{Eq:obj1}   \\
		\mathrm{s.t.} &LB\!\sum_{f=1}^{F}\!x_{m,f}^{L} \!\log\!\left( 1\!+\!\gamma_{m,f} p_{m,f}  \right) \!\geq \! R^L_m,   m\!=\!1,\!\ldots,\! M,\label{Eq:qos_embb}  \\
        & r^S_m \geq R^S_m, m=1,\ldots,M, \label{Eq:qos_urllc}  \\
		&\sum_{m=1}^M \left(x_{m,f}^L+ x_{m,f}^S\right) \leq 1, f=1,\ldots,F, \label{Eq:RB_con1}\\
		& x_{m,f}^L, x_{m,f}^S \!\in \!\{0,1\}, m\!=\!1,\!\ldots,\!M, f\!=\!1,\!\ldots,\!F, \! \label{Eq:RB_con2}\\
		& \sum_{f=1}^{F}\!\left(x_{m,f}^L\!+\!x_{m,f}^S\right)p_{m,f} \! \le \! P_{\max}, m\!=\!1,\!\ldots,\!M, \! \label{max_power} \\
		& p_{m,f}  \geq 0, m=1,\ldots,M, f= 1,\ldots,F,  \label{noneg_power}
	\end{align}
\end{subequations}
where for user~$m$, the variable $x_{m,f}^L\in\{0, 1\}$ indicates~whether RB$_f$ is assigned to the LBT, constraints in  \eqref{Eq:RB_con1} and \eqref{Eq:RB_con2} ensure that RB$_f$ can be assigned to at most one block of a user, constraints \eqref{Eq:qos_embb} and \eqref{Eq:qos_urllc} define the QoS requirements, \eqref{max_power} and \eqref{noneg_power} are the power constraints of users, and $P_{\max}$ denotes the maximal transmit power of each user.

Due to continuous power allocation and discrete RB allocation indicator, problem~\eqref{initial_problem} is a MINLP problem.
Thus, existing convex optimization techniques are not useful here.
Its global optimal solution can be obtained by  exhaustive searching, which, however, has a prohibitively high complexity. Hence, we resort to deep learning based methods to find a solution. Before delving into the learning methods, in the next section, we first develop two efficient numerical algorithms, 
which will serve as performance benchmarks for the subsequent  learning based~methods.


\section{Numerical Optimization for Resource Allocation}
In this section, we first consider a simplified version of problem~\eqref{initial_problem} involving a single user with both LBT and SBT data demands. We propose a hierarchical optimization algorithm to find the global optimal solution with low complexity. Then, based on this hierarchical method, a heuristic method is developed for multiuser scenario.

\subsection{Global Optimal Solution for Single-user Scenario} \label{subsection:Num_SU}
We consider the single-user scenario in this subsection, where $M=1$ and the subscript $m$ is omitted in problem~\eqref{initial_problem}.

While only a single user is considered, the problem is still complex to solve. We need to determine how many and which RBs should be allocated to LBT or SBT with how much power, aimed at minimizing the total number of occupied RBs, subject to the QoS requirements.

Considering that the LBT and SBT of a single user experience the same  channel conditions, an optimal solution to problem~\eqref{initial_problem} should select RBs with large channel gains in order to minimize the total number of required RBs. This suggests  a hierarchical algorithm. \textbf{In the outer loop}, the RBs are sorted in a descending order based on channel gains, and then the RBs with the highest channel gains are iteratively added into the occupied RBs set $\mathcal{N}$ until the QoS requirements are satisfied. \textbf{In the inner loop}, a feasibility problem is solved, determining whether the QoS requirements can be satisfied using the allocated RBs within~$\mathcal{N}$. The feasibility problem given $\mathcal{N}$ can be formulated~as
\begin{subequations}\label{inner_problem}
	\begin{align}
		{\!}\min_{x_{f}^{L},x_{f}^{S},p_{f}}  &  1  \label{inner_obj}\\
		\mathrm{s.t.}\ \   &\sum_{f\in\mathcal{N}}x_{f}^{L} \log\!\left( 1\!+\!\gamma_{f} p_{f}  \right) \!\geq \! \frac{R^L}{LB}, \\
		&\sum_{f\in\mathcal{N}}x_{f}^{S} \log \! \left( 1\!+\!\gamma_f p_f  \right)  \!-\! Q^{-1}\left(\epsilon\right)\sqrt{\frac{N^S}{ LB\tau}}\! \geq \! \frac{R^S}{LB},   \\
        & N^S = \sum_{f\in\mathcal{N}}x_{f}^{S}, \label{E:Nu} \\
		& x_f^L+ x_f^S = 1, f\in\mathcal{N}, \label{Eq:RB_con1_2}\\
		& x_f^L, x_f^S \in \{0,1\}, f\in\mathcal{N},  \label{Eq:RB_con2_2}\\
		& \sum_{f\in\mathcal{N}}p_f  \leq P_{\max},  \\
		& p_f > 0, f\in\mathcal{N}, \label{noneg_power_2}
	\end{align}
\end{subequations}
where the constraints come from the constraints of problem \eqref{initial_problem} by considering that all the RBs in set $\mathcal{N}$ are occupied in the inner loop.

This hierarchical algorithm sequentially increases the number of occupied RBs until all constraints in \eqref{initial_problem} are satisfied. Consequently, it is able to identify the minimum number of RBs required to satisfy all constraints, thereby achieving global optimality.

The feasibility of problem \eqref{inner_problem} can be determined by solving the following problem \eqref{maxrate_problem}, which maximizes the data rate of the LBT  while ensuring the QoS requirement of the SBT.
\begin{subequations}\label{maxrate_problem}
	\begin{align}
		{\!\!\!}\max_{x_{f}^{L},x_{f}^{S},p_{f}}  & LB \sum_{f\in\mathcal{N}}x_{f}^{L} \log \left( 1+\gamma_f p_f  \right) \label{E:obj111}  \\
		\mathrm{s.t.}  & \sum_{f\in\mathcal{N}}x_{f}^{S} \log \! \left( 1\!+\!\gamma_f p_f  \right)  \!-\! Q^{-1}\left(\epsilon\right)\sqrt{\frac{N^S}{ LB\tau}}\! = \! \frac{R^S}{LB}, \!\label{Eq:qos_urllc2}  \\
		& \sum_{f\in\mathcal{N}}p_f  =P_{\max},  \label{max_power_2} \\
        &\eqref{E:Nu},\eqref{Eq:RB_con1_2},\eqref{Eq:RB_con2_2},\eqref{noneg_power_2}.\notag
	\end{align}
\end{subequations}
Specifically, if the maximal value of the objective function in \eqref{E:obj111} is not smaller than $R^L$, then the RBs within $\mathcal{N}$ can meet the constraints of problem \eqref{inner_problem}, i.e., problem \eqref{inner_problem} is feasible. Otherwise, the RBs within $\mathcal{N}$ are insufficient, and problem \eqref{inner_problem} is infeasible.


\subsubsection{\underline{Optimal Power Allocation}}
We first study the optimal power allocation of problem \eqref{maxrate_problem}. By introducing Lagrangian multipliers $\eta$ and $\nu$ for constraints \eqref{Eq:qos_urllc2} and \eqref{max_power_2}, respectively, and  considering the positive power constraint in \eqref{noneg_power_2}, we can obtain the optimal power allocation from the Karush-Kuhn-Tucker (KKT) conditions (the details are omitted for the lack of space) as
\begin{equation}\label{power_forUe}
	p_f = \frac{x_f^L+x_f^S\eta}{\nu} -\frac{1}{\gamma_f}.
\end{equation}

The power allocation policy in \eqref{power_forUe} has a two water-level structure.
For the RBs allocated to the LBT, i.e., $\{\text{RB}_f|x_f^L=1, x_f^S=0,f\in\mathcal{N}\}$, \eqref{power_forUe} reduces to $p_f = \frac{1}{\nu} -\frac{1}{\gamma_f}$. The water-level is $\frac{1}{\nu}$, which should satisfy
\begin{align}
	\frac{1}{\nu} > \frac{1}{\gamma_{\min}^L} \label{water_level_positive2a}
\end{align}
to ensure the positive power constraint in \eqref{noneg_power_2}, where $\gamma _{\min}^{L}$ denotes the minimum of $\gamma_f$ for the RBs allocated to the LBT. Similarly, for the RBs allocated to the SBT, i.e., $\{\text{RB}_f|x_f^L=0, x_f^S=1,f\in\mathcal{N}\}$, \eqref{power_forUe} becomes $p_f = \frac{\eta}{\nu} -\frac{1}{\gamma_f}$. The water-level becomes $\frac{\eta}{\nu}$, which should satisfy
\begin{align}
	\frac{\eta}{\nu}>\frac{1}{\gamma_{\min}^S}, \label{water_level_positive2b}
\end{align}
where $\gamma _{\min}^{S}$ denotes the minimum of $\gamma_f$ for the RBs allocated to the SBT.

The  two water-levels can be derived by substituting \eqref{power_forUe} into \eqref{Eq:qos_urllc2} and \eqref{max_power_2}, which~are
\begin{subequations}\label{water_level_fin}
	\begin{align}
		\frac{1}{\nu} &=\frac{P_{\max}\!+\!\!\!\sum\limits_{f\in \mathcal{N}}\frac{1}{\gamma_f}\!-\!N^Se^{\frac{1}{N^S}(\frac{R^S}{LB}+Q^{-1}\left(\epsilon\right)\sqrt{\frac{N^S}{ LB\tau}}-\!\sum\limits_{f\in \mathcal{N}}\!\! \! x_f^S\log \gamma_f)}}{N^L},\\
		\frac{\eta}{\nu} &= e^{\frac{1}{N^S}(\frac{R^S}{LB}+Q^{-1}\left(\epsilon\right)\sqrt{\frac{N^S}{ LB\tau}}-\sum\limits_{f\in \mathcal{N}}x_f^S\log \gamma_f)},
	\end{align}
\end{subequations}
where $N^L=\sum_{f\in\mathcal{N}}x_{f}^{L}$ and $N^S=\sum_{f\in\mathcal{N}}x_{f}^{S}$ denote the total number of RBs allocated to the LBT and SBT, respectively, and  $N^L+N^S=N$ with $N$ denoting the total number of RBs in $\mathcal{N}$. It is clear that the water-levels depend on the RB allocation results $x_{f}^{L}$ and $x_{f}^{S}$.

\subsubsection{\underline{Optimal RB Allocation}}
By substituting the optimal power allocation in \eqref{power_forUe} and \eqref{water_level_fin} into problem \eqref{maxrate_problem} and considering the constraints in \eqref{water_level_positive2a} and \eqref{water_level_positive2b} for water-levels, we can eliminate $p_f$ and obtain an optimization problem only with respect to $x_{f}^{L}$ and $x_{f}^{S}$. The objective function in \eqref{E:obj111} can be derived as  \eqref{Eq:obj} (at the top of next page),
\begin{figure*}[htb]
\centering
\begin{align}\label{Eq:obj}
	&\ \ \ LB\sum\nolimits_{f\in \mathcal{N}}{x_{f}^{L}}\log  \Bigg( \frac{\gamma_f}{N^{L}} \Big(P_{\max}+\sum\nolimits_{f\in \mathcal{N}}\frac{1}{\gamma_f}- N^Se^{\frac{\frac{R^S}{LB}+Q^{-1}\left(\epsilon\right)\sqrt{\frac{N^S}{ LB\tau}}-\sum_{f\in \mathcal{N}}x_f^S\log \gamma_f}{N^{S}}} \Big) \Bigg)  \notag\\
	&=LB\log  \left( \frac{\prod_{f\in \mathcal{N}}{\gamma _f}}{\prod_{f\in \mathcal{N}}{x_f^S\gamma _f}}\left( \frac{1}{N^{L}} \left(P_{\max}+\sum\nolimits_{f\in \mathcal{N}}\frac{1}{\gamma_f}- N^Se^{\frac{\frac{R^S}{LB}+Q^{-1}\left(\epsilon\right)\sqrt{\frac{N^S}{ LB\tau}}-X}{N^{S}}} \right)\right) ^{N^{L}} \right) \\
	&=LB\left(\log\prod\nolimits_{f\in \mathcal{N}}{\gamma _f}+ Y-N^{L}\log N^{L}\right),\notag
\end{align}
\end{figure*}
where we define $X\triangleq\sum_{f\in \mathcal{N}}x_f^S\log{\gamma_f}$ and
$Y\triangleq -X+N^{L}\log( P_{\max}+\sum_{f\in \mathcal{N}}\frac{1}{\gamma_f}-N^{S}e^{\frac{\frac{R^S}{LB}+Q^{-1}\left(\epsilon\right)\sqrt{\frac{N^S}{ LB\tau}}-X}{N^{S}}})$ for brevity.

The objective function in \eqref{Eq:obj} is a complicated function of $x_f^S$ (or equivalently $x_f^L$ since $x_f^L+ x_f^S = 1$ in \eqref{Eq:RB_con1_2}), where the terms $X$, $Y$, $N^S$, and $N^L$ are all functions of $x_f^S$. Thus, it is difficult to directly optimize $x_f^S$ with \eqref{Eq:obj}. In the following we first optimize the RB allocation for any given $N^S$ and then find the best value of $N^S$. 

Given $N^S$, only $X$ and $Y$ in \eqref{Eq:obj} are related to $x_f^S$. Then, by omitting the irrelevant terms, we can rewrite problem \eqref{maxrate_problem} as the following problem
\begin{subequations}\label{P:optimalRB}
	\begin{align}
		\max_{x_f^S} \, & Y \\
		\mathrm{s.t.} & X\!>\! \underbrace{\begin{aligned}[t] &\frac{R^S}{LB}\!+\!Q^{-1}\!\left(\epsilon\right)\!\sqrt{\frac{N^S}{ LB\tau}}\! - \!
		N^S\log \Big( \frac{1}{N^S}\big( P_{\max}\\
		&+\sum\nolimits_{f\in \mathcal{N}}\frac{1}{\gamma_f}-\frac{N^L}{\gamma _{\min}^{L}} \big) \Big)\end{aligned}}_{\triangleq X^{lb}},\!\!\label{posi_e} \\
			& X \!<\! \underbrace{\frac{R^S}{LB}\!+\! Q^{-1}\left(\epsilon\right)\sqrt{\frac{N^S}{ LB\tau}}\!+\! N^S\log \gamma _{\min}^{S}}_{\triangleq X^{ub}},\!\label{posi_u}\\
	& x_f^S \in \{0,1\}, \label{discrete} \\
	&\eqref{E:Nu},\notag
	\end{align}
\end{subequations}
where $x^L_f$ is replaced by $1-x^S_f$ according to \eqref{Eq:RB_con1_2}, \eqref{posi_e} and \eqref{posi_u} come from the water-level constraints in \eqref{water_level_positive2a} and \eqref{water_level_positive2b}, and $X^{lb}$ and $X^{ub}$ denote the lower and upper bounds of $X$ in \eqref{posi_e} and \eqref{posi_u}, respectively.

In problem~\eqref{P:optimalRB}, the optimal RB allocation given $N^S$ is to choose $N^S$ RBs for the SBT, which can maximize $Y$ meanwhile satisfying constraints in \eqref{posi_e}$\sim$\eqref{discrete} and \eqref{E:Nu}.
To solve problem~\eqref{P:optimalRB}, let us first analyze the properties of the objective function $Y$.

\textbf{Property 1:} $Y$, defined in \eqref{Eq:obj}, is a concave function of $X$ since $\nabla^2_X Y < 0$. The maximum of $Y$ is achieved at $X^*$,
\begin{equation}\label{target_value}
	X^* \triangleq \frac{R^S}{LB}+Q^{-1}\left(\epsilon\right)\sqrt{\frac{N^S}{ LB\tau}}+N^{S}\log \frac{N}{P_{\max}+\sum_{f\in \mathcal{N}}\frac{1}{\gamma_f}},
\end{equation}
which can be obtained by letting $\nabla_{X} Y = 0$. Herein, $\nabla^2_X Y$ and $\nabla_X Y$ are the first and second derivatives of $Y$ with respect to $X$.

Note that the value of $X^*$ in \eqref{target_value} that maximizes $Y$ may fall outside the feasible region of $X$ in problem \eqref{P:optimalRB}. This is because constraints in \eqref{posi_e} and \eqref{posi_u} limit $X$ to the interval $(X^{lb},X^{ub})$ and constraints in  \eqref{discrete} and \eqref{E:Nu} restrict $X$ (i.e., $\sum_{f\in \mathcal{N}}x_f^S\log \gamma_f)$ to specific discrete values. Consequently, the optimal value of $X$ has the following property.

\vspace{0.2cm}
\textbf{Property 2:}
\begin{itemize}
	\item If $X^{ub}\leq X^*$, $Y$ increases with $X$ within $(X^{lb}, X^{ub})$. The optimal $X$ is the discrete value that is closest to $X^{ub}$ within $(X^{lb}, X^{ub})$.
	\item If $X^{lb}<X^*<X^{ub}$, $Y$ is concave for $X$. Let $X_1$ denote the discrete value that is closest to $X^*$ within $(X^{lb}, X^*]$ and $X_2$ denote the discrete value that is closest to $X^*$ within $(X^*, X^{ub})$. Then, the optimal $X$ is $X_1$ if $X_1$ leads to a larger value of $Y$ than $X_2$, or $X_2$ otherwise.
	\item If $X^*\leq X^{lb}$, $Y$ decreases with $X$ within $(X^{lb}, X^{ub})$. The optimal $X$ is the discrete value that is closest to $X^{lb}$ within $(X^{lb}, X^{ub})$.
\end{itemize}

Property 2 indicates that the optimal $X$ has three possible values, which are the discrete value closest to $X^{lb}$, $X^*$, or $X^{ub}$. The following Property 3 indicates which case should be selected. The detailed derivations can be found in Appendix~\ref{proof_he}.

\textbf{Property 3:}
\begin{align} \label{E:XLR5}
	\begin{cases}
		X^{lb} < X^*< X^{ub}, & \text{if}\ \gamma_{\min} > \psi,\\
		X^* \leq X^{lb}, & \text{if}\ \gamma_{\min} \leq \psi, \gamma_{\min}^{L} = \gamma_{\min},\\
		X^{ub} \leq X^*, & \text{if}\ \gamma_{\min} \leq \psi, \gamma_{\min}^{S} = \gamma_{\min},
	\end{cases}\notag
\end{align}
where $\psi\triangleq\frac{N}{P_{\max}+\sum_{f\in \mathcal{N}}\frac{1}{\gamma_f}}$, $\gamma_{\min}$ is the minimal channel gain of the RBs in $\mathcal{N}$, and the condition $\gamma_{\min}^{L} = \gamma_{\min}$ or $\gamma_{\min}^{S} = \gamma_{\min}$ means that the RB with the minimal channel gain is allocated to the LBT or SBT, respectively.


By combining Property 2 and Property 3, the optimal RB allocation can be determined by examining the three conditions listed in Property 3. For example, if $\gamma_{\min} \leq \psi$ and $\gamma_{\min}^{L} = \gamma_{\min}$, i.e., the second condition of Property 3  holds, then we have $X^* \leq X^{lb}$. This corresponds to Case 3 of Property 2, which indicates that the optimal RB allocation $x_f^S$ will make the optimal value of $X$ closest to $X^{lb}$ within $(X^{lb}, X^{ub})$. This leads to the following optimization problem
\vspace{-0.3cm}
\begin{subequations}\label{P:closest}
	\begin{align}
		\min_{x_f^S}\ \  & \sum\limits{_{f\in \mathcal{N}}}x_f^S\log \gamma_f - X^{lb}  \\
		\mathrm{s.t.}\ \ & X^{lb} < \sum\limits{_{f\in \mathcal{N}}}x_f^S\log \gamma_f < X^{ub},\label{posi_11}\\
		& \eqref{discrete},\eqref{E:Nu}.\notag
	\end{align}
\end{subequations}

Problem \eqref{P:closest} aims to select $N^S$ RBs from $\mathcal{N}$ such that $\sum_{f\in \mathcal{N}}x_f^S\log \gamma_f$ is closest to $X^{lb}$. This is a combinatorial optimization problem, which falls into the framework of ``subset-sum problems'' in the real-number domain. There exist efficient algorithms for solving the subset-sum problems, e.g., \cite{liu2016flsss}, which can be applied to solve problem \eqref{P:closest}.

We are now ready to present the whole algorithm to solve the original problem \eqref{initial_problem} where $M=1$.
The algorithm has a hierarchical structure. In the outer loop, we successively increase the number  of RBs $N$ with good channel gains  until the QoS requirements of both LBT and SBT are met. In the inner loop, we search the number of RBs $N^S$ allocated to the SBT  for a given $N$, and solve a series of subset-sum problems based on Property 2 and Property 3. We then obtain the optimal RB allocation policy for any specified $N^S$, with which the optimal power allocation can be obtained from~\eqref{power_forUe}.

The detailed algorithm is summarized in Algorithm \ref{AG:sig_approach}.
In the outer loop, $N$ is initialized by $N_{\min}$, which can be computed as follows. For any given number of RBs $N$, the maximal achievable rate of the user can be obtained by using the classic water-filling power allocation. If this maximal rate is lower than $R^L + R^S$, the available number of RBs $N$ is insufficient to support both the LBT and SBT. Accordingly, we set the minimal number of RBs as $N_{\min}+1$, ensuring the data rate not less than $R^L + R^S$.

\begin{algorithm}
	\caption{Optimal power and RB allocation for single-user scenario}
	\begin{algorithmic}[1] 	\label{AG:sig_approach}
		\REQUIRE Sort $\gamma_f, f=1,\ldots,F,$ in a descending order.
		\ENSURE $x_f^L$, $x_f^S$ and $p_f, f\in \mathcal{N}$.
		\STATE \textbf{Initialization:} $N=N_{\min}$ and $\mathcal{N}$ consists of $N$ best RBs.
		\REPEAT
		\STATE Add an unselected RB with the largest $\gamma_f$ to $\mathcal{N}$, and update $N\leftarrow N+1$.
		\FOR {$N^S=1,\ldots,N$}
		\IF {$\gamma_{\min}> \frac{N}{P_{\max}+\sum_{f\in \mathcal{N}}\frac{1}{\gamma_f}}$}
		\STATE Choose case 2 of Property 2.
		\STATE Obtain $x^S_f$ by solving the subset-sum problem with the  method in~\cite{liu2016flsss}.
		\STATE Compute the value of \eqref{Eq:obj}, denoted by $J(N^S)$.
		\ELSE
		\STATE \textbf{let} {$\gamma_{\min}^{L} = \gamma_{\min}$, i.e., $x^S_N = 0$} \textbf{do}
		\STATE\quad Choose case 3 of Property 2 and obtain $x^S_f$ by solving the subset-sum problem.
		\STATE\quad Compute the value of \eqref{Eq:obj}, denoted by $J_1$.
		
		\STATE \textbf{let} {$\gamma_{\min}^{S} = \gamma_{\min}$, i.e., $x^S_N = 1$} \textbf{do}
		\STATE \quad Choose case 1 of Property 2 and obtain $x^S_f$ by solving the subset-sum problem.
		\STATE \quad Compute the value of \eqref{Eq:obj}, denoted by $J_2$.
		\STATE \textbf{do} Compute $J(N^S) = \max\{J_1, J_2\}$.
		\ENDIF
		\ENDFOR
		\UNTIL{$\max\{J(N^S)\} \geq R^L$}
		\STATE Obtain $x_f^S, x_f^L=1-x_f^S, f\in \mathcal{N}$ corresponding to the maximum of $J(N^S)$.
		\STATE Compute $p_f, f\in \mathcal{N}$ with \eqref{power_forUe} and \eqref{water_level_fin}.
	\end{algorithmic}
\end{algorithm}

\subsection{Extended Algorithm for Multiuser Scenario} \label{subsection:Num_MU}
Extending the algorithm developed for the single-user scenario to the multiuser scenario is non-trivial. This is because in the multiuser case, it is no longer feasible to determine the occupied set of RBs for each user by simply sorting them based on the channel gains in the outer loop. Here, we propose a heuristic method in which users sequentially select RBs with the best channel conditions from the unselected RBs. This method is described as follows.
\begin{itemize}
  \item[i.] Every user, whose QoS requirements have not been fully satisfied, sequentially chooses an unselected RB with the largest channel gain, and adds this RB to the set of selected RBs.
  \item[ii.] With these selected RBs, each user employs the algorithm designed for the single-user scenario to optimize resource allocation and verifies  whether the QoS requirements for both LBT and SBT are satisfied.
  \item[iii.] Users who have met their QoS requirements will exit the iteration. The remaining users repeat step i and ii.
\end{itemize}

This method takes into account the QoS requirements of users by letting users exit the iteration once their QoS requirements are satisfied, meanwhile it ensures a fair allocation of RBs among users by the sequential selection process. Within each iteration, given the total number of RBs selected by each user, the RB and power allocation for each user can be solved by the proposed single-user algorithm.

Although the proposed numerical algorithms significantly reduce the computational complexity compared to exhaustive searching, these iterative algorithms are still unsuitable for real-time implementation. In the next section, we propose a deep learning based method to reduce inference complexity.

\section{Learning based Method for \\ Resource Allocation}
The primary difficulties of designing learning based method to solve problem \eqref{initial_problem} lie in two aspects. First, the policy to be learned involves discrete RB allocation, resulting in either vanishing or exploding gradient issues during the training of DNNs. Second, the SBT has stringent reliability requirements, requiring the learned policy to satisfy the QoS constraints with very high probability.

To tackle the difficulties, in this section we develop an adaptive smoothing method to facilitate the learning of discrete variables, and design a nonlinear penalty function that amplifies the  penalties for slight constraint violations.  

\subsection{Problem Transformation}
We begin with an equivalent transformation of problem~\eqref{initial_problem} with the following considerations. In problem \eqref{initial_problem}, we use $p_{m,f}$ to denote the power allocated by user $m$ to RB$_f$, and use the discrete variables $x_{m,f}^L$ and $x_{m,f}^S$ to denote whether RB$_f$ is allocated to the LBT or SBT for user~$m$.
However, the power allocation and RB allocation are related. For instance,
whether a RB is allocated to the LBT of a user is determined by whether the power allocated to the LBT of the user on the RB is positive or not. Thus, we can make an equivalent transformation of problem \eqref{initial_problem} that only involves power allocation variables. Then, RB allocations can be computed with corresponding powers. Such a  transformation can effectively reduce the solution space for learning by eliminating non-meaningful solutions, such as those cases where RBs are occupied yet no power is allocated or where  power is allocated to non-occupied RBs, e.g., $x^L_{m,f}=1$ but $p_{m,f}=0$. The transformation is achieved by defining new variables for power allocation,  ${p}^{L}_{m,f}$ and ${p}^{S}_{m,f}$, which denote the powers of user $m$ allocated to the LBT and SBT on RB$_f$, respectively.

To ensure that each RB is allocated to either the LBT or SBT of a single user, as constrained by \eqref{Eq:RB_con1}, at most one of power allocation variables ${p}^{s}_{m,f}$, for $s\in\{L, S\}$ and $m=1,\dots, M$, can be positive for RB$_f$, where $s\in\{L, S\}$  denotes LBT or SBT. To express this restriction, we introduce a piecewise maximal function $g{(\cdot)}$ as  
\begin{equation} \label{max_func2}
\begin{aligned}
&g^{s}_{m,f}\!\triangleq \! g{\left(p^{s}_{m,f}, \mathcal{P}^{s}_{m,f}\right)}\!=\!\begin{cases}
			p^{s}_{m,f}, &p^{s}_{m,f} \!>\! \max \mathcal{P}^{s}_{m,f},\\
			0, &\mathrm{else},
\end{cases}
\end{aligned}\!
\end{equation}
where $\mathcal{P}^{s}_{m,f} \!= \!\{p^{i}_{j,f} \mid  i \neq s \text{ or }j \neq m, i \in \{L, S\}, j=1,...,M\}$ denotes all the power allocation variables to RB$_f$ except for variable $p^{s}_{m,f}$. The function $g{(\cdot)}$ ensures that for $g_{m,f}^s, s\in\{L,S\}, m=1,\ldots,M$, at most one $g_{m,f}^s$ is positive while the rest are zero, thereby satisfying constraints in \eqref{Eq:RB_con1}.

With \eqref{max_func2}, the final power allocation of user $m$ on RB$_f$ can be obtained as $g^{s}_{m,f}$, and the final RB allocation decisions can be obtained as $\pmb{1}{(g^s_{m,f})}$, where $s\in\{L,S\}$ and $\pmb{1}(\cdot)$ is an indicator function that outputs 1 for positive inputs and 0 for zero input. Then, we can reformulate problem \eqref{initial_problem}~as
\begin{subequations}\label{convert_problem}
	\begin{align}
		&\!\!\!\!\!\!\!\!\!\! \min_{p^L_{m,f},p^S_{m,f}}  \sum_{f=1}^{F}{\pmb{1}}{\left({g}_{f}\right) } \label{Eq:obj1_c}   \\
		\mathrm{s.t.}
		& LB\!\sum_{f=1}^{F} \log\left( 1\!+\!\gamma_{m,f} {g}^L_{m,f}  \right) \!\geq \! R^L_m, m=1,\ldots,M,\label{Eq:qos_embb_c}  \\
		& LB\!\sum_{f=1}^{F} \log  \left( 1\!+\!\gamma_{m,f} {g}^S_{m,f} \right) \! - \!\sqrt{\frac{N^S_m LB}{\tau}}Q^{-1}\left(\epsilon_m\right) \!\geq \! R^S_m,\notag \\
		&\,\,\,\,\,\,\,\,\,\,\,\,\,\,\,\,\,\,\,\,\,\,\,\,\,\,\,\,\,\,\,\,\,\,\,\,\,\,\,\,\,\,\,\,\,\,\,\,\,\,\,\,\,\,\,\,\,\,\,\,\,\,\,\,\,\,\,\,\,\,\,\,\,\, m=1,\ldots,M, \label{Eq:qos_urllc_c}\\
		& \sum_{f=1}^{F}\left(p^L_{m,f} +p^S_{m,f} \right)  \le  P_{\max},m=1,\ldots,M,  \label{max_power2} \\
		& p^L_{m,f},p^S_{m,f}   \geq 0, f= 1,\ldots,F, m=1,\ldots, M, \label{noneg_power2}	
	\end{align}
\end{subequations}
where ${g}_{f}\triangleq\sum_{m=1}^{M}{g}^L_{m,f}+{g}^S_{m,f}$ denotes the total power~allocated to RB$_f$ by all users and $N^S_m= \sum_{f=1}^{F}{\pmb{1}}{({g}^S_{m,f})}$ denotes the total number of RBs allocated to the SBT of user~$m$.

Problem \eqref{convert_problem} only optimizes the power allocations $p_{m,f}^L$ and $p_{m,f}^S$, where the two discrete functions $g{(\cdot)}$ and $\pmb{1}(\cdot)$ ensure that the constraints in \eqref{Eq:RB_con1} and \eqref{Eq:RB_con2} are satisfied.
As analyzed before, the introduced discrete functions $g{(\cdot)}$ and $\pmb{1}(\cdot)$ can equivalently express the discrete RB allocation variables in problem \eqref{initial_problem}, making problem \eqref{initial_problem} and \eqref{convert_problem} have the same optimal solution. Although such a transformation still faces the gradient vanishing or explosion issues during training, it allows us to design smoothing methods to approximate the discrete functions.


Following the learning framework for constrained optimization proposed by \cite{Sun2019Learning}, problem \eqref{convert_problem} can be transformed into a primal-dual form by introducing Lagrangian multipliers to address constraints in \eqref{Eq:qos_embb_c} and \eqref{Eq:qos_urllc_c}. The primal-dual
problem is expressed as
\begin{subequations} \label{problem_func}
	\begin{align}
		\max_{\lambda^s_m}\min_{p^L_{m,f},p^S_{m,f}}& \sum_{f=1}^{F}\pmb{1}{\left({g}_{f}\right)} + \sum_{s\in\{L,S\}}\sum_{m=1}^M \lambda_m^s c_{m}^s \label{Lag_func}\\
		\mathrm{s.t.}\ \ & \lambda^s_m\geq 0, m=1,\ldots,M,s\in\{L,S\},\\
		& \eqref{max_power2},\eqref{noneg_power2},\notag
	\end{align}	
\end{subequations}
where $\lambda^L_m$ and $\lambda^S_m$ are the multipliers associated with constraints in \eqref{Eq:qos_embb_c} and \eqref{Eq:qos_urllc_c}, respectively, and $c_{m}^L \triangleq R^L_m-LB\sum_{f=1}^{F} \log( 1+\gamma_{m,f} {g}^L_{m,f})$ and $c_{m}^S\triangleq R^S_m-LB\sum_{f=1}^{F} \log  ( 1+\gamma_{m,f} {g}_{m,f}^S )  +\sqrt{N^S_mLB/\tau}Q^{-1}\!\left(\epsilon_m\right)$ are defined based on the QoS constraints in \eqref{Eq:qos_embb_c} and \eqref{Eq:qos_urllc_c}. 

\subsection{Adaptive Smoothing Functions}
\begin{figure}
	\centering
	\includegraphics[width=0.8\linewidth]{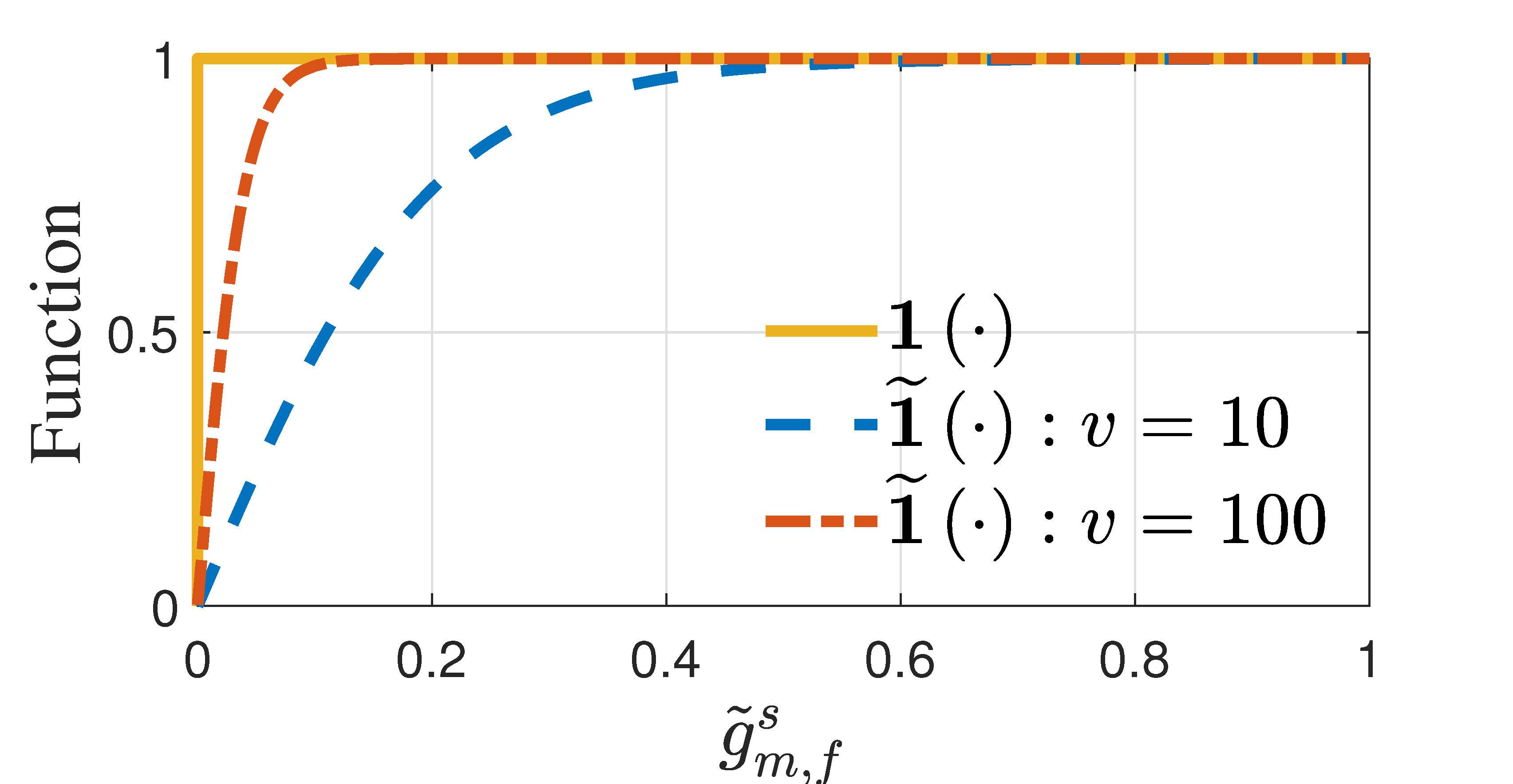}
	\caption{Impact of different values of $v$ on smoothing function $\tilde{\pmb{1}}\left(\cdot\right)$.}
	\label{fig:smooth}
\end{figure}

The piecewise maximal function $g{(\cdot)}$ and the indicator function $\pmb{1}(\cdot)$ involved in problem \eqref{problem_func} are non-differentiable. These two functions pose obstacles for the gradient-based training of DNNs since their gradients are either zero or infinity, causing DNNs converging to poor solutions or not converging at all. To deal with the non-differentiable functions, we develop two smoothing functions, denoted as $\tilde{g}\left(\cdot\right)$ and $\tilde{\pmb{1}}\left(\cdot\right)$, to approximate the original functions $g{(\cdot)}$ and $\pmb{1}(\cdot)$, respectively. $\tilde{g}\left(\cdot\right)$ is a variation of SoftMax function and $\tilde{\pmb{1}}\left(\cdot\right)$ is a variation of Tanh function, which are defined as
\begin{subequations}\label{smooth_func}
		\begin{equation}\label{smooth_pic_E}
		   \tilde{g}^s_{m,f}\triangleq \tilde{g}{\left( p^s_{m,f}, \mathcal{P}^s_{m,f},u\right)}=\frac{p^s_{m,f}}{\sum_{i\in\{L,S\}}\sum_{j=1}^{M}e^{u(p^i_{j,f}-p^s_{m,f})}},
	\end{equation}
	\begin{equation}\label{smooth_indi_E}
	\tilde{\pmb{1}}\left( \tilde{g}^s_{m,f},v\right)=  \frac{2}{1+e^{-v \tilde{g}^s_{m,f}}} -1,
	\end{equation}
\end{subequations}
where $u, v\geq0$ are the smoothing parameters. The values of $u$ and $v$ affect the approximation errors. It is not difficult to find that when $u$ and $v$ are infinite, the approximation errors become zeros. Meanwhile, $u$ and $v$ also affect the smoothness, i.e., gradient, of the approximation functions, as shown in Fig.~\ref{fig:smooth}. 
Increasing the smoothing parameters can improve the approximation accuracy but simultaneously reduce the smoothness, and vice versa. 

Existing methods either adopt fixed smoothing parameters or simply
adjust them as the number of iterations increases, such as the annealing technique. These methods assume a one-size-fits-all parameter setting that is intended to be effective when the variable in a smoothing function
takes different  values, which however is difficult if not impossible. For example, considering the smoothed indicator function $\tilde{\pmb{1}}(\tilde{g}^s_{m,f},v)$ in \eqref{smooth_indi_E}, as shown in Fig. \ref{fig:smooth}, using a larger smoothing parameter results in the gradient $\nabla_{\tilde{g}^s_{m,f}}\tilde{\pmb{1}}( \tilde{g}^s_{m,f},v)$ approaching zero for large values of $\tilde{g}^s_{m,f}$, thereby resulting in gradient vanishing issues. Conversely, a smaller smoothing parameter introduces large approximation errors for small values of $\tilde{g}^s_{m,f}$. During the training process, the input of the smoothing functions may take different values among iterations, making it difficult to find an appropriate smoothing parameter.

We propose to adaptively adjust the smoothing parameters $v$ and $u$ in  $\tilde{\pmb{1}}\left(\cdot\right)$ and $\tilde{g}{\left( \cdot\right)}$. Since it is difficult to find a universal parameter suitable for all variable values, we let the values of $v$ and $u$ change with the variable values. Inspired by the annealing technique, in the early iterations of training, smaller smoothing parameters are preferred to avoid gradient vanishing or exploding, facilitating faster convergence, while as the number of iterations increases, reducing the approximation errors becomes more crucial. Thus, we let the values of $v$ and $u$ also adjust with the number of iterations.

Next we elaborate the design of the adaptive smoothing parameters for the two functions.
%

\subsubsection{\underline{Smoothing indicator function}}
With the smoothing functions in \eqref{smooth_indi_E}, the indicator functions used in problem \eqref{problem_func}, i.e.,
$\pmb{1}{({g}_{f})}$ and ${\pmb{1}}{({g}^S_{m,f})}$, can be approximated by $\tilde{\pmb{1}}{(\tilde{g}_f,v)}$ and $\tilde{\pmb{1}}{(\tilde{g}_{m,f}^S,v)}$, respectively, where $\tilde{g}_f = \sum_{m=1}^{M}\tilde{g}^L_{m,f}+\tilde{g}^S_{m,f}$. For ease of discussion, we take $\tilde{\pmb{1}}{(\tilde{g},v)}$ as an example to elaborate the parameter adaptation, where subscript~$f$~is~omitted.


In order to prevent DNNs from converging to a poor solution or failing to converge due to the gradients that are either too large or too small, we adjust the smoothing parameter $v$ for each value of $\tilde{g}$ to make the absolute value of gradient  $|\nabla_{\tilde{g}}\tilde{\pmb{1}}{(\tilde{g},v)}|$ approach a predetermined value (called gradient requirement in the sequel). Moreover, this gradient requirement is not fixed, which evolves during training. At the start, a larger gradient can accelerate network convergence, whereas a reduced gradient can decrease approximation errors as training progresses. Therefore, we let the gradient requirement decrease as the number of  iterations increases. 

Now given an iteration during the training process, for which the gradient requirement has a value of $V$, let us examine how to determine the corresponding value of $v$. Taking into account both the approximation error and gradient requirement, the problem of finding the smoothing parameter $v$ can be formulated as minimizing the approximation error under the gradient requirement, which is
\begin{subequations}\label{P:smooth_indi}
	\begin{align}
	\min_{v\geq 0} &\left|\tilde{\pmb{1}}{(\tilde{g},v)}-\pmb{1}{(\tilde{g})}\right|\\
	\mathrm{s.t.}&|\nabla_{\tilde{g}}\tilde{\pmb{1}}{(\tilde{g},v)}| = V.\label{P:smooth_indi_c}
	\end{align}
\end{subequations}


As proven in Appendix \ref{proof_gradient}, problem \eqref{P:smooth_indi} is feasible if $\frac{2\zeta e^{-\zeta}}{\tilde{g}(1+e^{-\zeta})^2} > V$ holds, in which case the optimal solution of $v$ can be obtained by using bisection search on the equation $|\nabla_{\tilde{g}}\tilde{\pmb{1}}{(\tilde{g},v)}|\!=\!V$, where $\zeta\!\approx\!1.54$ is a constant. If problem~\eqref{P:smooth_indi} is infeasible, i.e., the required gradient $V$ cannot be achieved, then we find the smoothing parameter $v$ that can maximize $|\nabla_{\tilde{g}}\tilde{\pmb{1}}{(\tilde{g},v)}|$, whose solution is $\frac{\zeta}{\tilde{g}}$ as derived in Appendix~\ref{proof_gradient}.

To summarize, the smoothing parameter $v$ is computed for every iteration based on the current value of $\tilde{g}$. 
The obtained $v$ ensures that the absolute value of gradient $|\nabla_{\tilde{g}}\tilde{\pmb{1}}{(\tilde{g},v)}|$ is close to the gradient requirement $V$, which can avoid the vanishing or exploding of gradient. The value of $V$ decreases as the number of iterations increases, resulting in gradually reduced approximation errors.
Similarly, the smoothing parameter for function $\tilde{\pmb{1}}{(\tilde{g}_m^S,v)}$ can also be adaptively adjusted during the training process, considering both the number of iterations and the specific value of $\tilde{g}_m^S$.

\subsubsection{\underline{Smoothing piecewise maximal function}}
The smoothing piecewise maximal function $\tilde{g}_{m,f}^s$ in \eqref{smooth_pic_E} is a multivariate function of the power allocation vector $\pmb{p}_f=[p^L_{1,f},\cdots,p^L_{M,f},p^S_{1,f},\cdots,p^S_{M,f}]\in \mathbb{R}^{1\times 2M}$. We assign different smoothing parameters for each element of $\pmb{p}_f$, denoted as $\pmb{u}=[u^L_1,\cdots,u^L_{M},u^S_1,\cdots,u^S_{M}]\in \mathbb{R}^{1\times 2M}$. Then, the expression of $\tilde{g}_{m,f}^s$ in  \eqref{smooth_pic_E} can be redefined with $\pmb{u}$ as
\begin{equation}  \label{max_func2bb}
 \tilde{g}{\left(p_{m,f}^s, \mathcal{P}_{m,f}^s,\pmb{u}\right)}=\frac{p_{m,f}^s}{\sum_{i\in\{L,S\}}\sum_{j=1}^{M}e^{u^i_j(p^i_{j,f}-p^s_{m,f})}}.
\end{equation}
For notational simplicity, the subscript~$f$ is omitted in the sequel of this subsection.

The design of the smoothing parameters $\pmb{u}$ should take into account both the approximation error and gradient requirement. 
With \eqref{max_func2} and \eqref{max_func2bb}, the absolute value of approximation error $\left|\tilde{g}_{m}^s-g_{m}^s\right|$ can be obtained as
\begin{equation}\label{obj_two}
\begin{cases}
p_{m}^s-\frac{p_{m}^s}{\sum_{i\in\{L,S\}}\sum_{j=1}^{M}e^{u^i_j(p^i_{j}-p^s_{m})}},& p_{m}^s > \max\{\mathcal{P}_{m}^s\},\\
\frac{p_{m}^s}{\sum_{i\in\{L,S\}}\sum_{j=1}^{M}e^{u^i_j(p^i_{j}-p^s_{m})}},& p_{m}^s <\max\{\mathcal{P}_{m}^s\}.	
\end{cases}
\end{equation}
The gradient $\nabla_{\pmb{p}}\tilde{g}^s_{m}$ is a vector, and we employ the $\ell_1$ norm of $\nabla_{\pmb{p}}\tilde{g}^s_{m}$ to measure the gradient magnitude, i.e., $\lVert\nabla_{\pmb{p}}\tilde{g}^s_{m}\rVert_1=\sum_{i\in\{L,S\}}\sum_{j=1}^{M}|\nabla_{p^i_{j}}\tilde{g}^s_{m}|$, which can be derived~as
\begin{align}\label{E:gradient}
  \frac{1}{\sum\limits_{i\in\{L,S\}}\sum\limits_{j=1}^{M}e^{u^i_j(p^i_{j}-p^s_{m})}}  \!+\!\frac{2p^s_{m}\!\!\!\!\sum\limits_{\substack{i\in\{L,S\},j=1,\ldots,M\\i \neq s \text{ or } j \neq m}}\!\!\!\! u^i_je^{u^i_j(p^i_{j}-p^s_{m})}}{\big(\sum\limits_{i\in\{L,S\}}\sum\limits_{j=1}^{M}e^{u^i_j(p^i_{j}-p^s_{m})}\big)^2}.
\end{align}

For the case $p^s_{m} > \max\{\mathcal{P}^s_{m}\}$, we can find from \eqref{obj_two} that the approximation error approaches zero if $\sum_{i\in\{L,S\}}\sum_{j=1}^{M}e^{u^i_j(p^i_{j}-p^s_{m})}$ approaches one. Since $e^{u^i_j(p^i_{j}-p^s_{m})}=1$ when $i=s$ and $j=m$, we know that $\sum_{i\in\{L,S\}}\sum_{j=1}^{M}e^{u^i_j(p^i_{j}-p^s_{m})}$ approaching one means that $e^{u^i_j(p^i_{j}-p^s_{m})}$ should approach zero for any $i\neq s$ or $j\neq m$. When this happens, it can be found from \eqref{E:gradient} that the gradient magnitude $\lVert\nabla_{\pmb{p}}\tilde{g}^s_{m}\rVert_1$ approaches one, thus avoiding the gradient vanishing or exploding. To achieve such a desired result with zero approximation error, we can choose $u^s_m$ as $0$ for ${p}^s_{m}$ and choose $u^i_j$ as $\frac{-\rho}{{p}_{j}^i-{p}_{m}^s}$ for ${p}^i_{j}\neq {p}^s_{m}$, where $\rho$ is a large positive constant.

For the case $p^s_{m} < \max\{\mathcal{P}^s_{m}\}$, we observe from \eqref{obj_two} that the approximation error approaches zero as  $\sum_{i\in\{L,S\}}\sum_{j=1}^{M}e^{u^i_j(p^i_{j}-p^s_{m})}\to\infty$. However, this will lead to a vanishing gradient as shown in \eqref{E:gradient}, necessitating the optimization of the smoothing parameter $\pmb{u}$. The problem of optimizing $\pmb{u}$ can be formulated similar to \eqref{P:smooth_indi}~as
\begin{subequations}\label{P:smooth_pm}
	\begin{align}
	\min_{\pmb{u} \succeq0}\ & \left|\tilde{g}^s_{m}-g^s_{m}\right| \label{E:objsmooth}\\
    \mathrm{s.t.}\ &\lVert\nabla_{\pmb{p}}\tilde{g}^s_{m}\rVert_1=\bar{V},\label{P:smooth_pm1}
	\end{align}
\end{subequations}
where $\bar{V}$ is the gradient requirement varying with the number of  iterations, and $\succeq$ is the element-wise~inequality.

Problem \eqref{P:smooth_pm} has a complicated objective function, making it difficult to obtain its global optimal solution. Yet, as analyzed in Appendix~\ref{proof_gradient2}, the second term of $\lVert\nabla_{\pmb{p}}\tilde{g}^s_{m}\rVert_1$ in \eqref{E:gradient} is very small when the objective function \eqref{E:objsmooth} is minimized. By omitting this term, we obtain an approximate solution for the smoothing parameters as derived in Appendix~\ref{proof_gradient2}, which is
\begin{align} \label{E:opt-u0}
  u_{j}^i\!=\!	\begin{cases}
		\frac{1}{{p}_{j}^i-{p}_{m}^s}\log {\frac{1/\bar{V}-2M+|\mathcal{G}|}{|\mathcal{G}|}},& \frac{1}{1+|\mathcal{G}|}\geq \bar{V}, {p}^i_{j}>{p}^s_{m},\\
        \frac{-\rho}{{p}_{j}^i-{p}_{m}^s}, & \frac{1}{1+|\mathcal{G}|}< \bar{V}, {p}^i_{j}<{p}^s_{m},\\
		0, &\mathrm{else},
	\end{cases}
\end{align}
where $|\mathcal{G}|$ is the size of the set $\mathcal{G}\triangleq\{{p}^i_{j}|{p}^i_{j}>{p}^s_{m},i\in\{L,S\},j=1,\ldots,M\}$.

\subsection{Nonlinear Penalty to Constraint Violation}\label{subs:nonlinear}
A constrained optimization problem can be solved through learning by transforming it into a primal-dual form~\cite{Eisen2019Learning,Sun2019Learning}. The introduced Lagrangian multipliers are learned by DNNs, referred to as multiplier DNNs.

The training of multiplier DNNs is related with the values of  $c^s_m$ in \eqref{problem_func}, since the gradients of the objective function in \eqref{Lag_func} with respect to multipliers $\lambda_m^s$ is $c^s_m$.
The sign of $c^s_m$ determines the update direction of $\lambda_m^s$, while the magnitude of $c^s_m$ affects the update step size. If a constraint is nearly met, the magnitude of $c^s_m$ will be very small, resulting in a slow update of the multiplier DNN. As a result, the  multiplier is hard to increase rapidly enough to penalize the violated constraints, thereby causing a high probability of constraint violation.

The SBT has a very stringent reliability requirement, which is reflected by the very low decoding error probability $\epsilon_m$ in the considered problem.
When learning based methods are employed, the learned policy may not always satisfy constraints, and hence the probability of constraint violation, denoted as $\hat{\epsilon}_m$, also affects the achieved reliability. The total error probability can be expressed as
\begin{equation}\label{error}
 1-(1-\hat{\epsilon}_m)(1-\epsilon_m)\approx \hat{\epsilon}_m+\epsilon_m,
\end{equation}
where the approximation holds since both $\hat{\epsilon}_m$ and $\epsilon_m$ are very small. It indicates that the learning based method should ensure a very low level of constraint violations.

To reduce constraint violation, the penalty for minor violations of constraints should be increased. We address this by introducing a nonlinear function $q\left(\cdot\right)$ on $c_m^s$ in \eqref{problem_func} to amplify the penalty of constraint violations. With $q\left(\cdot\right)$ and the smoothing functions, the Lagrangian function in \eqref{Lag_func} can be expressed~as
\begin{equation} \label{E:obj}
	\begin{aligned} \sum_{f=1}^{F}\tilde{\pmb{1}}\left(\tilde{g}_{f},v\right)+ \sum_{s\in\{L,S\}}\sum_{m=1}^M \lambda^s_m q\left(c_{m}^s\right).
	\end{aligned}
\end{equation}

The nonlinear function $q\left(\cdot\right)$ is designed as
\begin{equation}\label{g_func}
	q\left(c_m^s\right)=
	\begin{cases}
		\kappa\big(\frac{2}{1+e^{-w c_m^s}}-1\big), & c_m^s> 0,\\
		-\min\big\{\frac{\lambda_m^s}{2},1\big\}, & c_m^s\leq 0.
	\end{cases}
\end{equation}
The principle of the design and its impact on the penalty of constraint violation are explained as follows.

When $c_m^s>0$, i.e., a constraint violation happens, $q\left(c_m^s\right)$ is designed as $\kappa\big(\frac{2}{1+e^{-w c_m^s}}-1\big)$, which is a smoothed indicator function $\tilde{\pmb{1}}(c_m^s)$ as defined in \eqref{smooth_indi_E}, where $\kappa$ is a constant.
Such a $q\left(c_m^s\right)$ causes a sharp increase of the violation value from $c_m^s$ to $\kappa(\frac{2}{1+e^{-w c_m^s}}-1)\rightarrow \kappa$. The smoothing parameter $w$ is computed with the previously proposed adaptive smoothing parameter method, which can avoid the gradient vanishing and gradient exploding caused by introducing the nonlinear function $q\left(\cdot\right)$. On the other hand, when the constraint is satisfied, $q\left(c_m^s\right)$ is designed as $-\min\{\frac{\lambda_m^s}{2},1\}$. For small $\lambda_m^s$, say less than two,
$q\left(\cdot\right)$ equals to $-\frac{\lambda_m^s}{2}$. Upon substituting it into \eqref{E:obj}, we can find that the gradient with respect to the multiplier is $-\lambda^s_m$, i.e., we reduce the multipliers based on their current values. To prevent an over reduction of multipliers, we restrict that the decreasing rate of multipliers does not exceed one, which is realized by the minimal function $\min\{\frac{\lambda_m^s}{2},1\}$.

\emph{Remark:} When applying the nonlinear function $q\left(c_m^s\right)$ to $c_m^s$ in the Lagrangian function \eqref{E:obj}, we have transformed the constraint from $c_m^s\leq0$ to $q\left(c_m^s\right)\leq 0$.  Such a transformation is equivalent, because it can be found from \eqref{g_func} that  $q\left(c_m^s\right)\leq 0$ always implies $c_m^s\leq0$. 

\subsection{Proposed Learning Method} \label{S:learn_net}
With the previously developed two smoothing functions and the nonlinear penalty function, we are ready to present our learning method.

According to the proof in \cite{Sun2019Learning},
the primal-dual formulation \eqref{problem_func} can be equivalently transformed as the following functional optimization problem
\begin{subequations} \label{E:functional1}
	\begin{align}
		\max_{\lambda^s_m  \!\left(\pmb{\gamma}\right)}\min_{p^s_{m,f}\!\left(\pmb{\gamma}\right)}&\,	\mathbb{E}\Big\{\! \sum_{f=1}^{F}\!{\pmb{1}}\left({g}_{f},
	v\right)\! +\!\! \!\!\!\!\sum_{s\in\{L,S\}}\!\sum_{m=1}^M \! \lambda^s_m  \!\left(\pmb{\gamma}\right) c_{m}^s\!\Big\}\!\!\\
		\mathrm{s.t.}\ \ \  & \!\!\!\!\!\sum_{s\in\{L,S\}}\!\sum_{f=1}^{F}p^s_{m,f}\!\left(\pmb{\gamma}\right) \! \le \! P_{\max},m\!=\!1,\!\ldots,\!M, \!\!\! \\
		& \pmb{p}\left(\pmb{\gamma}\right), \pmb{\lambda}^s\left(\pmb{\gamma}\right)\succeq 0,s\in\{L,S\},
	\end{align}	
\end{subequations}
where $p^s_{m,f}(\pmb{\gamma})$ and $\lambda^s_m  (\pmb{\gamma})$ are the  functions of the channel gains $\pmb{\gamma}\triangleq\left[\gamma_{1,1},\cdots,{\gamma}_{M,F}\right]\in\mathbb{R}^{1\times MF}$, defining $\lambda^s_m  (\pmb{\gamma})$ as a function is necessary because  the constraints in \eqref{Eq:qos_embb_c} and \eqref{Eq:qos_urllc_c} should
be satisfied for all the possible values of $\pmb{\gamma}$,  $\mathbb{E}\{\cdot\}$ denotes the expectation taken over~$\pmb{\gamma}$, $\pmb{p}\left(\pmb{\gamma}\right)=\left[p^L_{1,1}(\pmb{\gamma}),\cdots,p^S_{M,F}(\pmb{\gamma})\right]\in\mathbb{R}^{1\times 2MF}$, and  $\pmb{\lambda}^s(\pmb{\gamma})=\left[\lambda^s_1(\pmb{\gamma}),\cdots, \lambda^s_M(\pmb{\gamma})\right]\in\mathbb{R}^{1\times M}$.
\vspace{0.1cm}

We employ DNNs to parameterize the functions $\pmb{p}\left(\pmb{\gamma}\right)$ and $\pmb{\lambda}^s(\pmb{\gamma})$. Instead of learning directly from  channel~gains  $\pmb{\gamma}$, we consider a sorted version of  $\pmb{\gamma}$. As pointed out by  \cite{sun2022improving}, if  the inputs and outputs of a DNN are permutation equivariant along a  dimension, then sorting the inputs along that dimension can significantly reduce the  required training samples.


In our problem, the input channel gains and output power allocations are permutation equivariant along the dimension of RB. Specifically, reordering the channel gains associated with the RBs leads to a corresponding reordering in the optimal power allocation. For the single-user scenario, sorting the channel gains is straightforward since each RB has a single channel gain. For the multiuser scenario, however, sorting becomes difficult because each RB has multiple channel gains from different users. We present a heuristic sorting method for the multiuser scenario, which is shown to be effective by simulation results later. The method is inspired by the multiuser RB selection strategy given in Sec.~\ref{subsection:Num_MU}, which operates as follows. First, the RB on which the first user has the largest channel gain among $F$ RBs is selected as the first RB. Next, the RB on which the second user has the largest channel gain among the remaining $F-1$ RBs is selected as the second RB. This sequential process continues until all RBs are sorted. 

After the sorting of the channel gains $\pmb{\gamma}$, the power allocation policy, i.e., the mapping from the sorted $\pmb{\gamma}$ to the corresponding power allocation $\pmb{p}$, is parameterized by a \emph{policy DNN} $\pmb{p}\left(\pmb{\gamma};\pmb{\theta}_{p}\right)$, where $\pmb{\gamma}$ is the input and  $\pmb{\theta}_{p}$ is the trainable parameter. The functions   $\pmb{\lambda}^s(\pmb{\gamma})$,  $s\in\{L,S\}$ are parameterized by two \emph{multiplier DNNs}, $\pmb{\lambda}^s\left(\pmb{\gamma}; \pmb{\theta}_{s}\right)$,  $s\in\{L,S\}$ respectively, where $\pmb{\theta}_{s}$ is the trainable parameter. With the parameterization, problem \eqref{E:functional1} is converted into the optimization with respect to the trainable parameters $\pmb{\theta}_{s}$ and $\pmb{\theta}_{p}$. Further considering the previously proposed two smoothing functions and the nonlinear penalty function, they are obtained from the following optimization~problem
\begin{subequations} \label{learn_problem}
	\begin{align}
		\max_{\pmb{\theta}_{s}}\min_{\pmb{\theta}_{p}}&\,	\mathbb{E}\Big\{\! \sum_{f=1}^{F}\!\tilde{\pmb{1}}\left(\tilde{g}_{f},
	v\right)\! +\!\! \!\!\!\!\sum_{s\in\{L,S\}}\!\sum_{m=1}^M \! \lambda^s_m  \!\left(\pmb{\gamma}; \pmb{\theta}_{s}\right) q\big(c_{m}^s \big)\!\Big\}\!\! \label{E:obj-learn}\\
		\mathrm{s.t.}\ \ \  & \!\!\!\!\!\sum_{s\in\{L,S\}}\!\sum_{f=1}^{F}p^s_{m,f}\!\left(\pmb{\gamma};\pmb{\theta}_{p}\right) \! \le \! P_{\max},m\!=\!1,\!\ldots,\!M, \!\!\! \label{normal_con} \\
		& \pmb{p}\left(\pmb{\gamma};\pmb{\theta}_{p}\right), \pmb{\lambda}^s\left(\pmb{\gamma}; \pmb{\theta}_{s}\right)\succeq 0,s\in\{L,S\}.\label{posi_con}
	\end{align}	
\end{subequations}

The training of the policy DNN and two multiplier DNNs is implemented by the primal-dual stochastic gradient method. The primal parameters $\pmb{\theta}_{p}$ and the dual parameters $\pmb{\theta}_L$ and $\pmb{\theta}_S$ are iteratively updated along the descent and ascent directions of the sample-averaged gradients of the loss function $\mathcal{L}$, respectively, where $\mathcal{L} = \sum_{f=1}^{F}\!\tilde{\pmb{1}}\left(\tilde{g}_{f},
	v\right)\! +\!\sum_{s\in\{L,S\}}\!\sum_{m=1}^M \! \lambda^s_m  \!\left(\pmb{\gamma}; \pmb{\theta}_{s}\right) q\big(c_{m}^s \big)$.
Constraints in \eqref{normal_con} and \eqref{posi_con} can be satisfied by properly choosing the activation function at the output layer of both the policy DNN and multiplier DNNs. 


\begin{figure}
	\centering
	\includegraphics[width=1\linewidth]{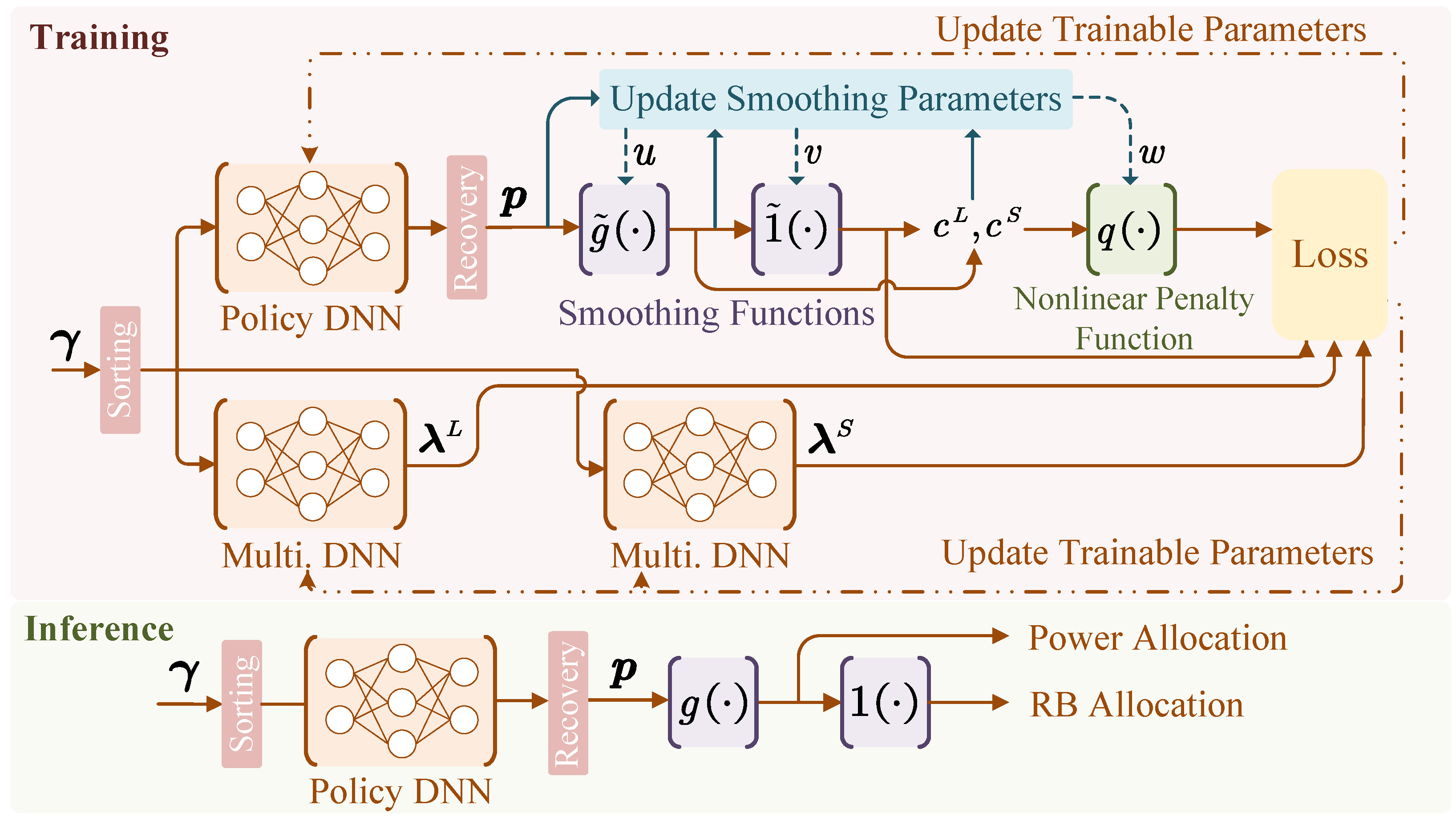}
	\caption{Training and inference processes of the policy DNN and multiplier DNNs (denoted as multi. DNNs).}
	\label{fig:Network_model}
\end{figure}

In Fig. \ref{fig:Network_model}, we illustrate the training process for the policy and  multiplier DNNs as well as the inference process with the well-trained policy DNN, i.e., offline training and online inference.
During each iteration, we first sort the samples in a batch based on the sorting method presented in this subsection. Next, we compute and update the smoothing parameters in two smoothing functions and a nonlinear penalty function based on the methods proposed in the previous subsections. Subsequently, the trainable parameters of the three DNNs are updated using the loss function computed with the updated smoothing parameters. For the inference process, the smoothing functions are no longer used. The power allocation can be obtained by inputting the outputs of the well-trained policy DNN into the piecewise maximal  function~\eqref{max_func2}, while the RB allocation can be obtained by inputting the power allocation into the indicator function $\pmb{1}(\cdot)$.

\section{Simulation Results}
In this section, we evaluate the performance of the proposed methods by simulations.

\subsection{Simulation Setup}
Consider an uplink OFDMA system, where the BS is equipped with $N_r=64$ receive antennas, the subcarrier spacing is $B=30$~kHz, each RB contains $L=12$ subcarriers, the total number of RBs is $F=40$, and the duration of a time slot is $\tau = 0.5$~ms. The channels follow the Saleh-Valenzuela model, where the number of scattered paths is 10 and the angle of arrival follows a uniform distribution within $[-\pi/3,\pi/3]$. The path loss is modeled as $35.3+37.6\log _{10}{d}$ in dB, where $d$ is the distance between $M=2$ users and the BS, which is set to $150$~m. The maximal transmit power of each user is $23$~dBm, single-sided noise spectral density is $-174$~dBm/Hz, the penetration coefficient is $20$ dB, the noise figure of the BS receiver is $5$ dB, and the interference margin is $2$ dB~\cite{GPP_NF}.
The minimal rate requirements for the SBT and LBT are $R^S = 512$~kbps and $R^L = 6$~Mbps, respectively, unless otherwise specified. The required error probability for the SBT is set to  $10^{-5}$~\cite{GPP913}.
To achieve such a reliability requirement, we set the decoding error probability as $10^{-5}$ for the proposed numerical methods. For the learning based methods, we set both the decoding error probability and the constraint violation probability as $5 \times 10^{-6}$, which can ensure a final error probability of $10^{-5}$ according to \eqref{error}.

To implement the proposed learning based method, we build three fully-connected DNNs as the policy and multiplier DNNs, respectively. The three DNNs have three fully-connected hidden layers, each with 1000 neurons  in the single-user scenario and with 2000 neurons in the multiuser one. Their input layers have $MF$ neurons, and their output layers have $2MF$, $M$ and $M$ neurons, respectively.
All DNNs use Softplus as the activation function of the hidden layers. Policy DNN uses  ReLU  as the activation function of output layer. For each user, the allocated power is first normalized and then multiplied by $P_{\max}$ to ensure constraint in \eqref{normal_con}. Multiplier DNNs use  Softplus  as the activation function of their output layers. Adam is used to adjust the learning rate during training, and the initial learning rate is  $5\times {10}^{-3}$.
The Batch normalization is used to normalize the inputs of each layer, and the batch size is  $400$. The training and testing sets consist of ${10}^{7}$ and ${10}^{6}$ samples $\pmb{\gamma}$, respectively, which are generated with the channel, path loss and noise models specified in the previous paragraph.
The standardization method is applied to the input data to improve training stability and convergence. 
In the adaptive parameter methods, the gradient requirement $V$ increases linearly from $10$ to $80$ during the first $5\times 10^4$ iterations with warmup strategy \cite{he2016deep} and then decreases linearly from $80$ to $20$, while $\bar{V}$ decreases linearly within $[10^{-3},10^{-5}]$. In the nonlinear penalty function, the  gradient requirement is set to $0.4$, and the parameter $\kappa$ increases exponentially from $0.5$ to $20$. These setups are used in the sequel unless otherwise~specified.

\subsection{Performance Evaluation}
We compare the numerical method proposed in Sec.~\ref{subsection:Num_SU} for the single-user scenario (with legend ``\emph{Prop-Num-SU}"), the numerical method with heuristic multiuser RB selection proposed in Sec.~\ref{subsection:Num_MU} for the multiuser scenario (with legend ``\emph{Prop-Num-MU}"), and the proposed learning based method (with legend ``\emph{Prop-Learn}") with the following baselines.
\begin{enumerate}
	\item \emph{Exhaustive Search}: This method searches through all possible combinations of RB allocation, for each of which the optimal power allocation is computed. Then, the combination achieving the best~performance is selected.
	\item \emph{Fixed Parameter}: This method uses fixed smoothing parameters to approximate the piecewise maximal function and indicator function, as considered in \cite{Fang2020Stochastic}, where $v=50$ and $u=200$. It is simulated to validate the advantage of the proposed adaptive smoothing method.
	\item \emph{Annealing}: This method gradually adjusts the smoothing parameters using the annealing technique during  training~\cite{jang2016categorical}, where $v$ and $u$ increase linearly within $[50,400]$ and $[200,500]$, respectively. It is also used for showing the advantage of the adaptive smoothing~method.
	\item \emph{Default-Constr}: This method is the standard primal-dual based learning method without introducing the nonlinear penalty functions~\cite{Eisen2019Learning}. The fine-tuned learning rate of policy DNN is linearly decayed from $5\times {10}^{-3}$ to $5\times {10}^{-4}$. It is simulated to validate the advantage  of the nonlinear penalty of constraint violation.
	\item \emph{Incr-Require}: This is also the standard primal-dual based learning method but with increased rate requirements for LBT and increased decoding error probability requirements for SBT during training~\cite{Sun2019Learning}, which leads to a conservative policy to reduce the fraction of violated constraints. The fine-tuned learning rate of policy DNN is linearly decayed from $5\times {10}^{-3}$ to $5\times {10}^{-4}$.
	\item \emph{Fixed-Multi}: This method uses a fixed weight to penalize violated constraints as considered in \cite{liang2019towards}. It is also employed to show the advantage  of the nonlinear penalty.
	\item \emph{Unsorted Samples}: This is a modified version of \emph{Prop-Learn}, which does not sort training samples. The comparison with it can reflect the impact of sorting~samples.
\end{enumerate}

In what follows, the performance metrics are averaged over the samples in the testing set, except for that in Fig. \ref{fig:Fig4}.

\begin{figure} [h]
	\centering
	\subfloat[Average number of occupied RBs.]{\label{fig:Fig1_1}
		\includegraphics[width=0.8\linewidth]{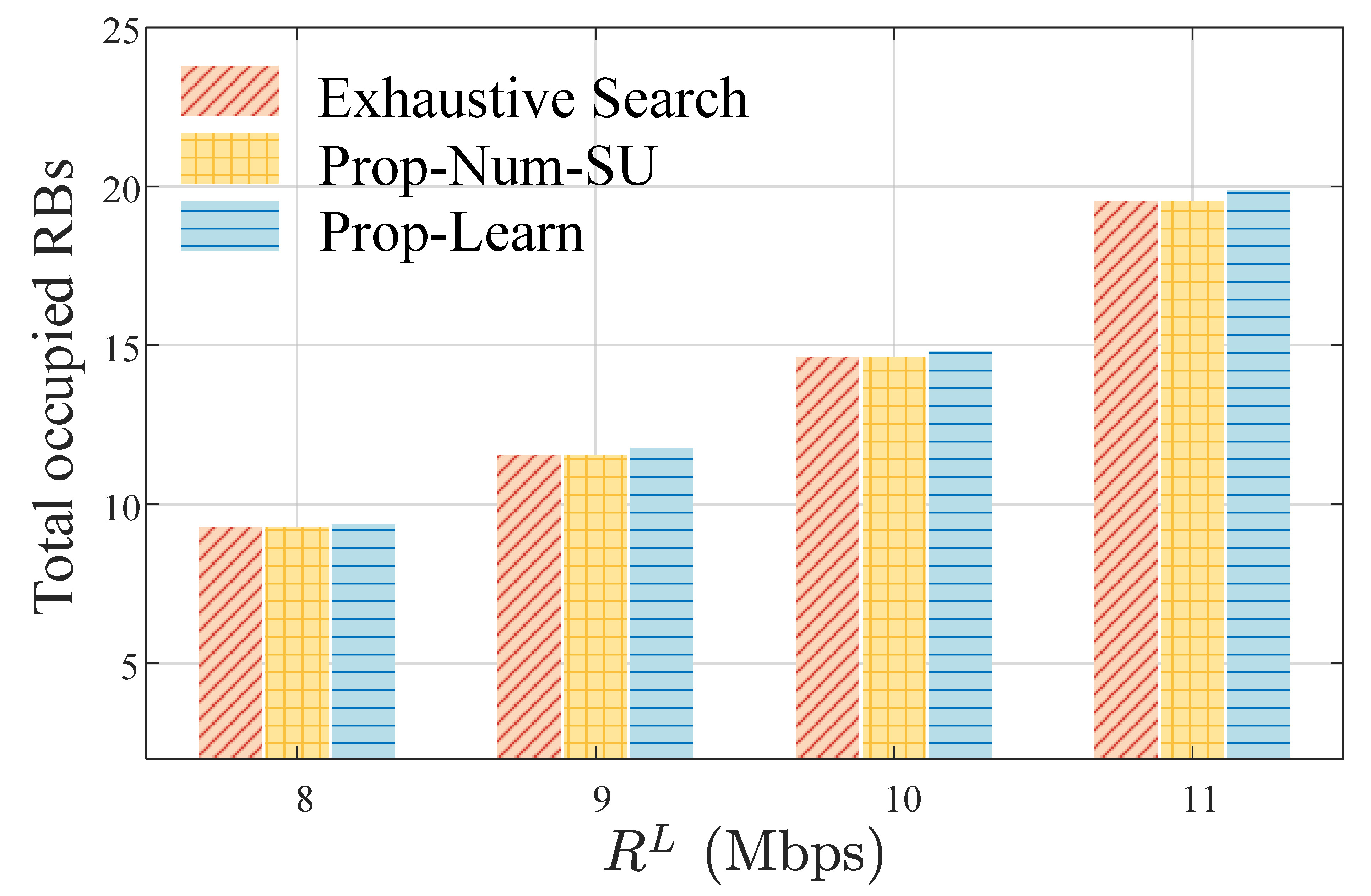}
	} 	
	\quad
	\subfloat[Running time.]{ \label{fig:Fig1_2}
		\includegraphics[width=0.8\linewidth]{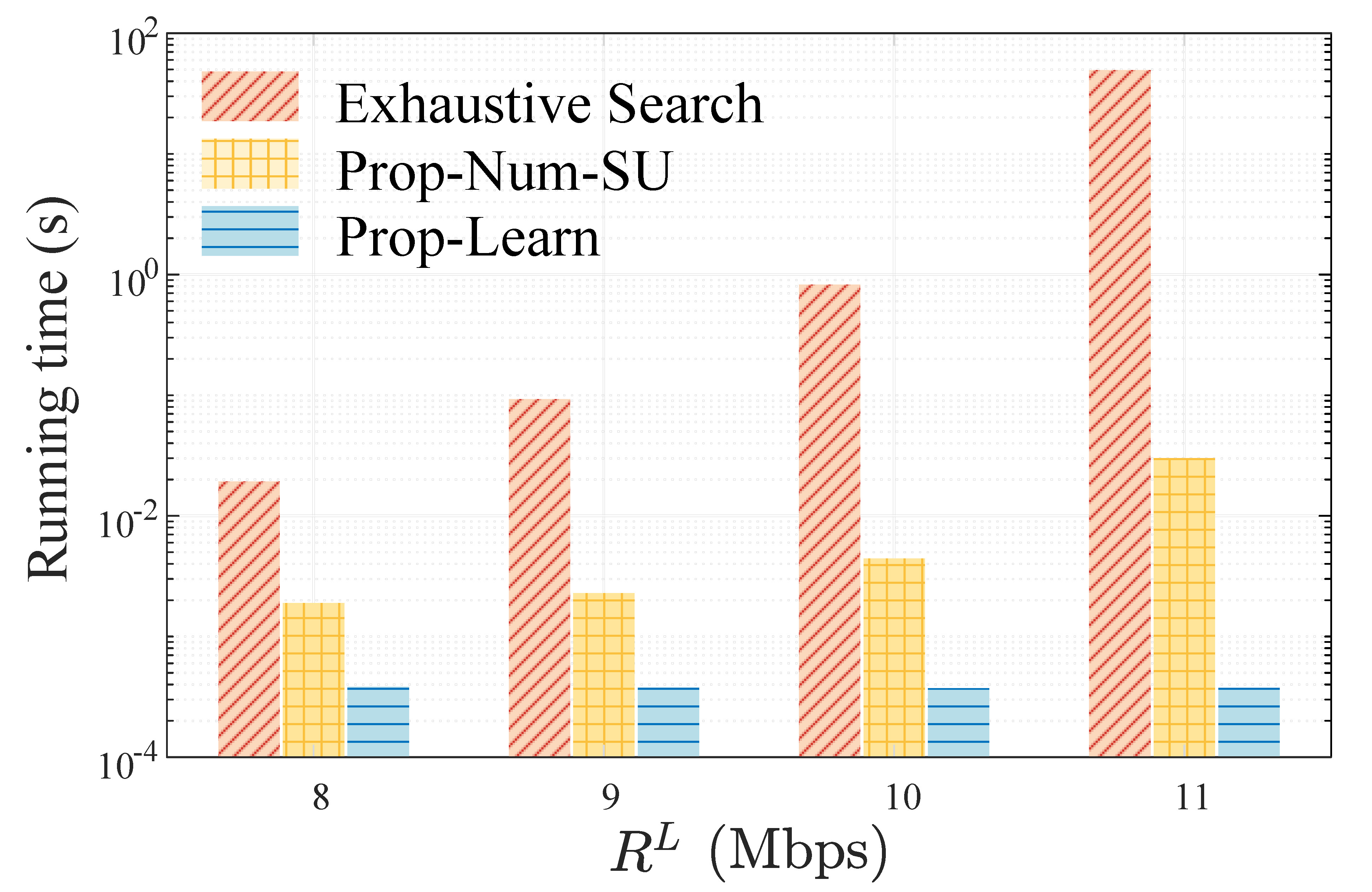}
	}
	\caption{Performance comparison with exhaustive searching.
	}
	\label{fig:Fig1}
\end{figure}

\subsubsection{\underline{Performance and complexity of the numerical method}}
Figure~\ref{fig:Fig1} compares the performance and complexity of \emph{Prop-Num-SU} and \emph{Prop-Learn} with \emph{Exhaustive Search} for the single-user scenario.
It is shown by Fig.~\ref{fig:Fig1_1} that \emph{Prop-Num-SU} achieves the same minimal number of occupied RBs as \emph{Exhaustive Search}, and \emph{Prop-Learn} performs very close to the numerical algorithms. In  Fig.~\ref{fig:Fig1_2}, the running time of the three methods is evaluated, which is averaged over random channels on a personal computer. Compared to \emph{Exhaustive Search}, \emph{Prop-Num-SU} and \emph{Prop-Learn} can reduce the complexity by several orders of magnitude. For \emph{Prop-Learn}, the running time is the inference time of a well-trained policy DNN, which remains nearly constant as $R^L$ increases.


\subsubsection{\underline{Learning with sorted samples}}
\begin{figure} 
	\centering
	\subfloat[Average number of occupied RBs.]{\label{fig:Fig2_1}
		\includegraphics[width=0.8\linewidth]{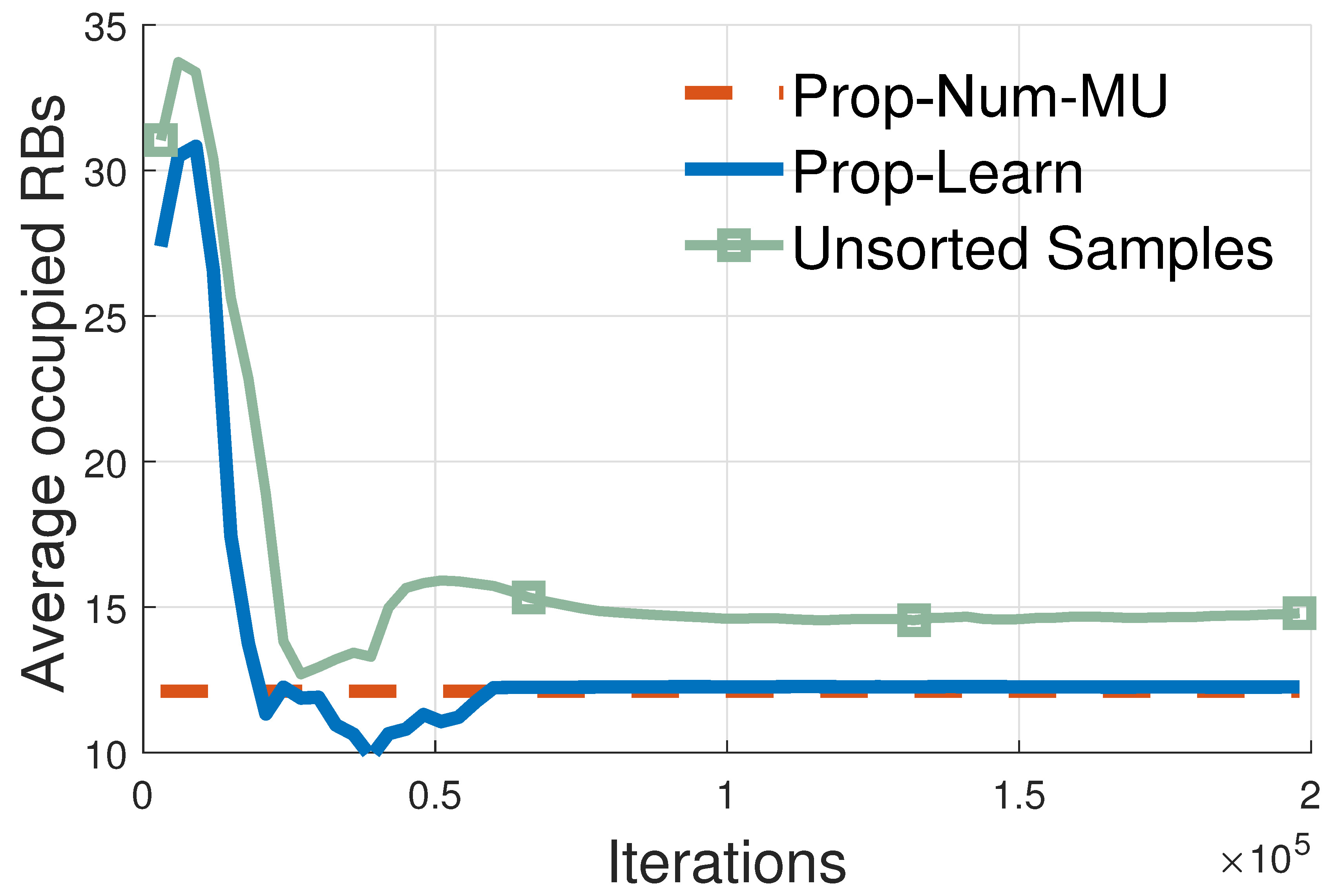}
	} 	
	\quad
	\subfloat[Constraint violation fraction.]{ \label{fig:Fig2_2}
		\includegraphics[width=0.8\linewidth]{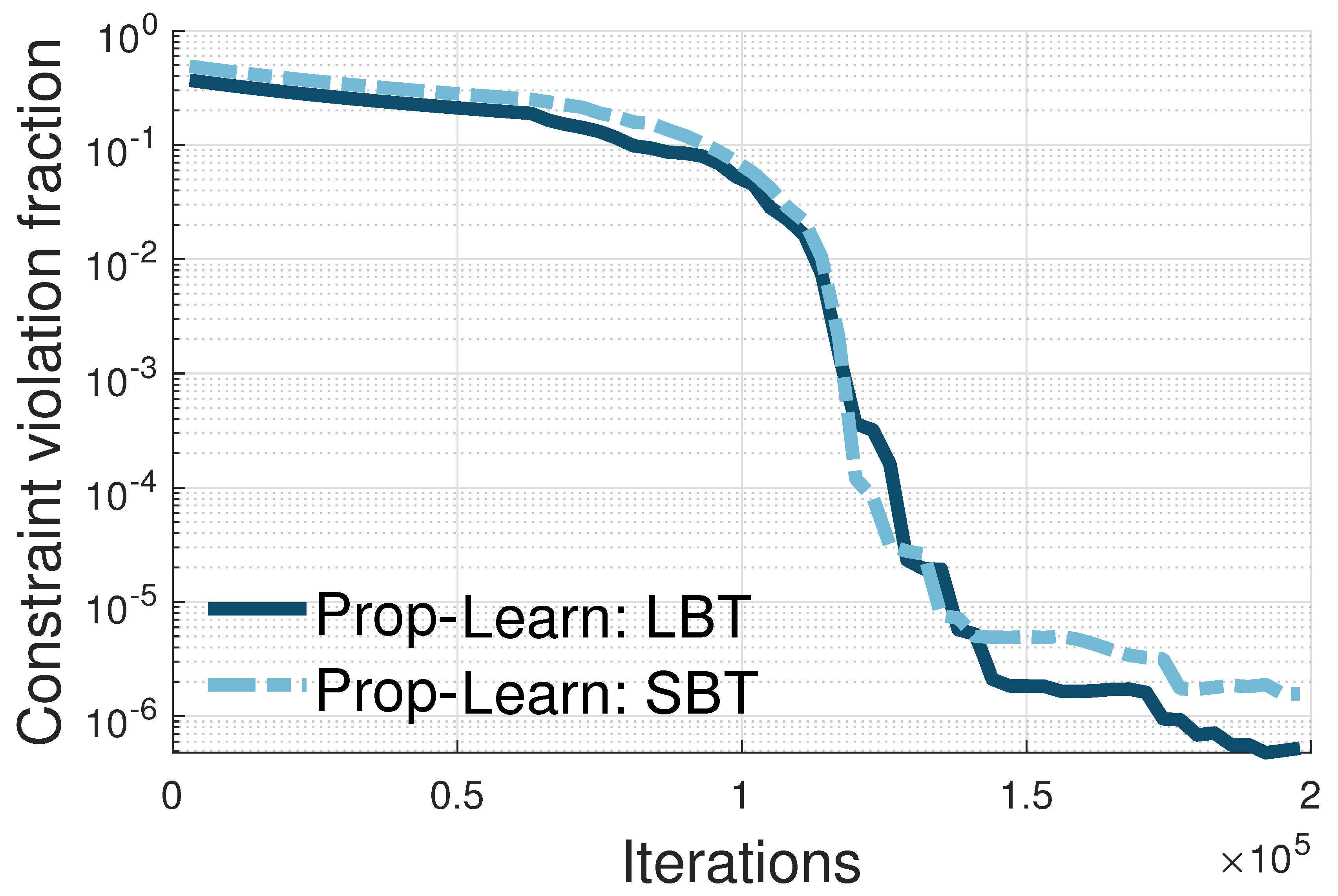}
	}
	\caption{Learning curves for performance and constraint violation (averaged over 20 iterations).}
	\label{fig:Fig2}
	\vspace{-0.3cm}
\end{figure}

Figure~\ref{fig:Fig2} depicts the learning curves for the total number of occupied RBs and the fraction of constraint violations. As shown in Fig.~\ref{fig:Fig2_1}, compared to the learning method with unsorted samples, \emph{Prop-Learn} with  sorting can  reduce the total number of occupied RBs meanwhile accelerating the convergence. \emph{Prop-Learn} performs close to \emph{Prop-Num-MU}. From Fig.~\ref{fig:Fig2_2}, we can find that the well-trained \emph{Prop-Learn} can achieve a very low fraction of constraint violations for both LBT and SBT below the target level of  $5 \times 10^{-6}$.

\renewcommand{\arraystretch}{1.3}
\begin{table}
	\centering
	\caption{Performance and constraint violation comparison}
	\label{table1}
	\resizebox{\columnwidth}{!}{
		\begin{tabular}{|cl|l|ll|}
			\hline\hline
			\multicolumn{2}{c|}{\multirow{2}{*}{}}                                 & \multicolumn{1}{c|}{\multirow{2}{*}{\shortstack{Average number\\ of occupied RBs}}} & \multicolumn{2}{c}{Constraint violation fraction}     \\ \cline{4-5}
			\multicolumn{2}{c|}{}                                                  & \multicolumn{1}{c|}{}                                     & \multicolumn{1}{c|}{LBT} & \multicolumn{1}{c}{SBT} \\ \hline
			\multicolumn{2}{c|}{\emph{Prop-Num-MU}}                                     & \multicolumn{1}{c|}{12.12}                                    & \multicolumn{1}{c|}{0} & \multicolumn{1}{c}{0}     \\ \hline
			\multicolumn{2}{c|}{\emph{Prop-Learn}}                             & \multicolumn{1}{c|}{12.26}                                    & \multicolumn{1}{c|}{$5.24\times10^{-7}$}  & \multicolumn{1}{c}{$1.57\times10^{-6}$}   \\ \hline
			\multicolumn{2}{c|}{\emph{Fixed Parameter}}                      &  \multicolumn{1}{c|}{13.64}                                                          & \multicolumn{1}{c|}{$4.38\times10^{-3}$}     &      \multicolumn{1}{c}{$7.91\times10^{-3}$}                      \\ \hline
			\multicolumn{2}{c|}{\emph{Annealing}}                      &    \multicolumn{1}{c|}{13.17}    & \multicolumn{1}{c|}{0.014}     & \multicolumn{1}{c}{0.024}    \\ \hline
			\multicolumn{2}{c|}{\emph{Default-Constr}}                      &  \multicolumn{1}{c|}{12.11}                                                          & \multicolumn{1}{c|}{0.21}     &      \multicolumn{1}{c}{0.22}                      \\ \hline
			\multicolumn{2}{c|}{\emph{Incr-Require}}                      &    \multicolumn{1}{c|}{13.49}    & \multicolumn{1}{c|}{0.013}     & \multicolumn{1}{c}{0.021}    \\ \hline
			\multicolumn{1}{c|}{\multirow{2}{*}{\shortstack{\emph{Fixed-}\\ \emph{Multi}}
			}} & $\lambda^s_m\!=\!10^2$  &         \multicolumn{1}{c|}{13.62}   & \multicolumn{1}{c|}{$1.29\times10^{-5}$}     &          \multicolumn{1}{c}{0.020}                    \\ \cline{2-5}
			\multicolumn{1}{c|}{}                                           & $\lambda^s_m\!=\!10^4$ &           \multicolumn{1}{c|}{13.79}                                               & \multicolumn{1}{c|}{$2.25\times10^{-7}$}     &
			\multicolumn{1}{c}{$1.16\times10^{-6}$}                       \\ \hline\hline
		\end{tabular}
	}
\end{table}

\subsubsection{\underline{Adaptive parameter method}} Table~\ref{table1} shows the impact of the proposed adaptive smoothing method, where the performance metrics in Table~\ref{table1} are averaged across different random seeds. Compared to \emph{Fixed Parameter} and \emph{Annealing}, \emph{Prop-Learn} occupies fewer RBs while achieving a lower fraction of constraint violations. Both \emph{Fixed Parameter} and \emph{Annealing} are difficult to obtain appropriate smoothing parameters to prevent gradient vanishing or exploding, leading to inefficient RB usage and large constraint violations.

\begin{figure} 
	\centering
	\subfloat[LBT.]{\label{fig:Fig4_1}
		\includegraphics[width=0.8\linewidth]{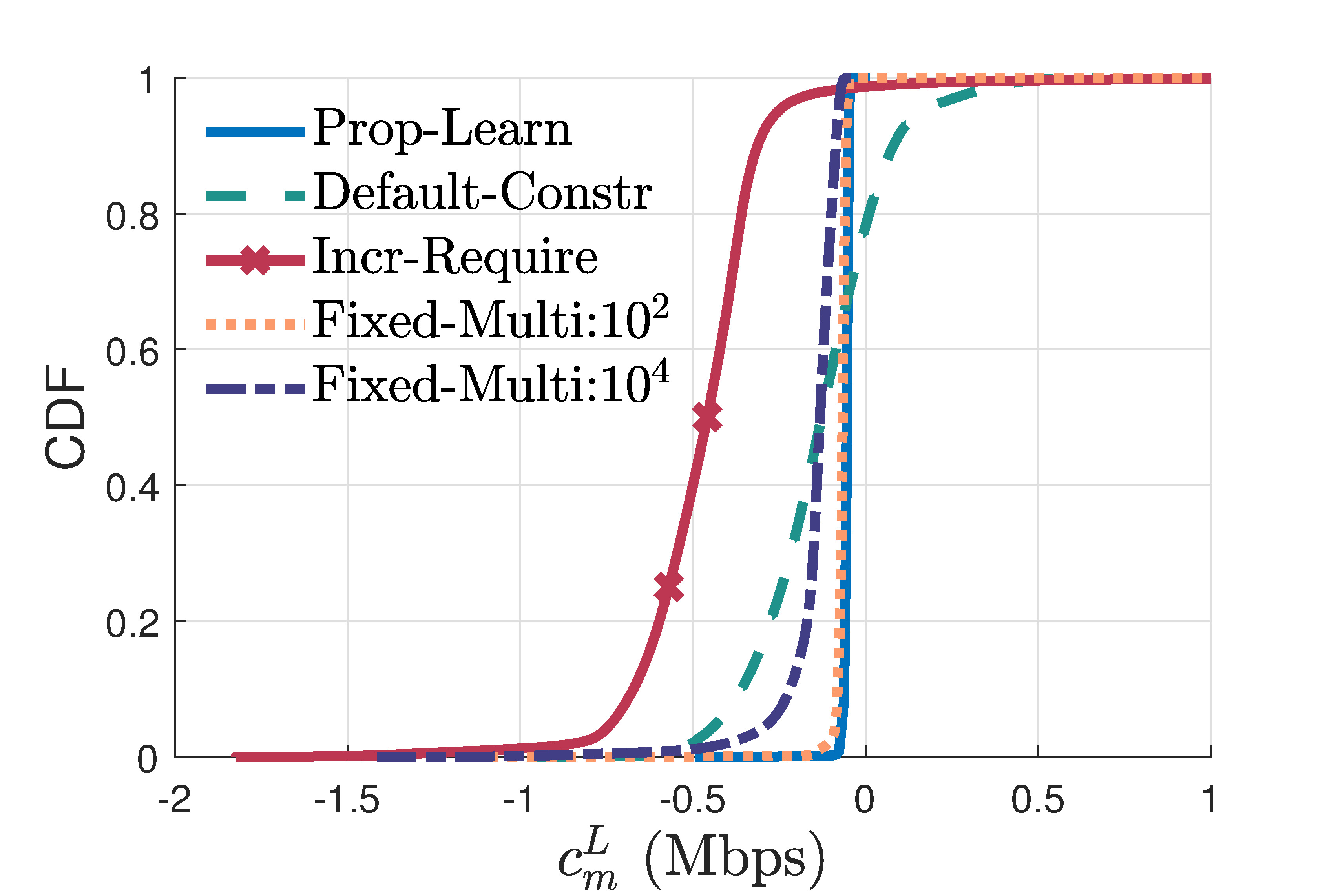}
	} 	
	 \quad
	\hspace{-2mm}
	\subfloat[SBT.]{ \label{fig:Fig4_2}
		\includegraphics[width=0.8\linewidth]{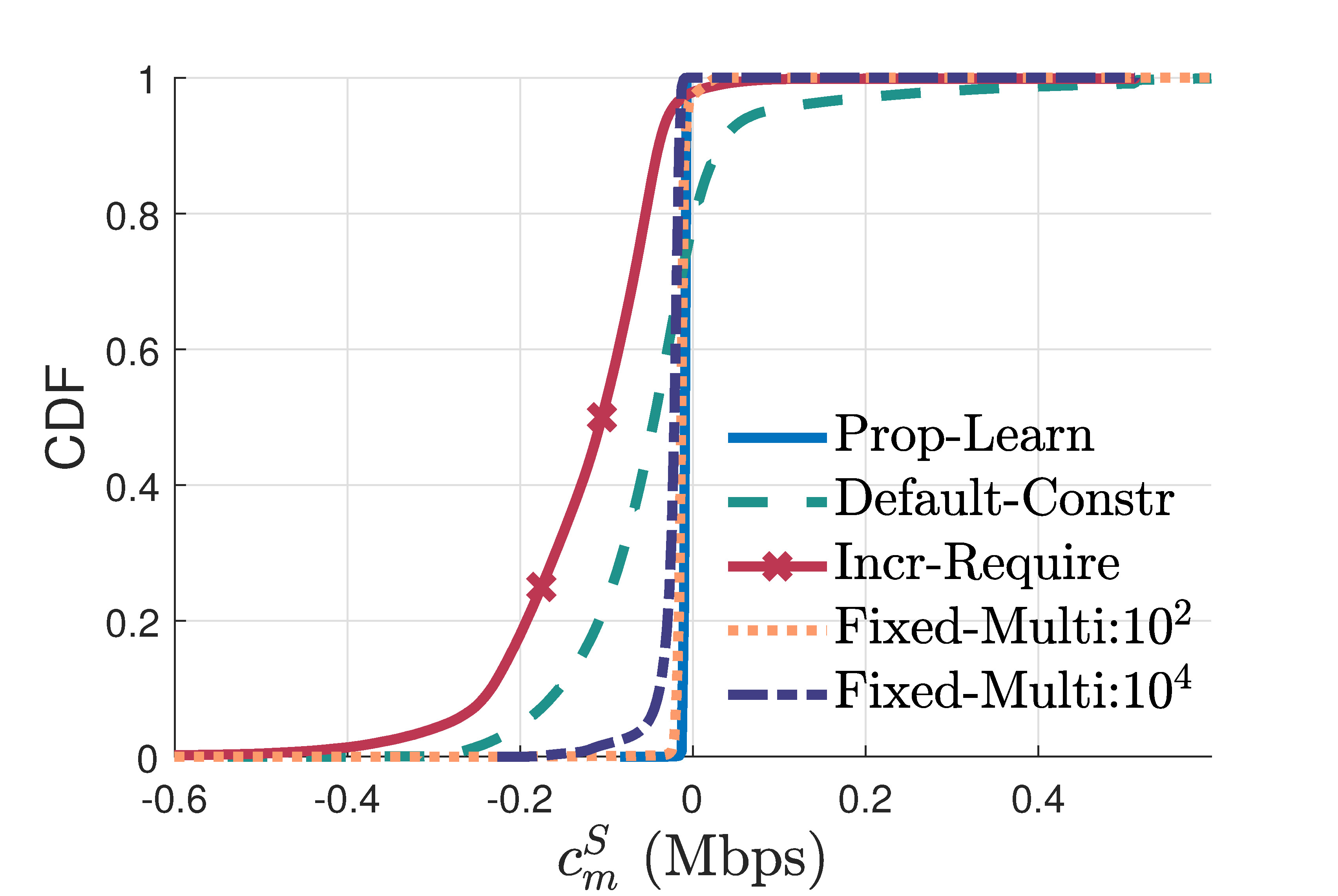}
	}
	\caption{CDF of $c^L_m$ and $c^S_m$ for two blocklength transmissions.
	}
	\label{fig:Fig4}
	\vspace{-0.2cm}
\end{figure}

\subsubsection{\underline{Nonlinear penalty function for QoS constraints}} We~demonstrate the impact of the designed nonlinear penalty function in Table~\ref{table1} and Fig.~\ref{fig:Fig4}. Fig.~\ref{fig:Fig4} shows the cumulative distribution function (CDF) of $c^L_m$ and $c^S_m$ in the testing set. A negative value of $c^L_m$ or $c^S_m$ means an over-satisfaction of QoS constraints, whereas a  positive value indicates a violation.

Without employing the nonlinear penalty function, \emph{Default-Constr} exhibits a high fraction ($>20\%$) of constraint violations. Through increasing the rate requirements of LBT by 5\% and  decoding error probability requirements of SBT by $10^{-8}$, \emph{Incr-Require} only reduces the constraint violations to approximately $1.3\%$ or $2.1\%$ at the cost of increasing more than one RB. For \emph{Fixed-Multi}, different values of the multipliers result in different learning outcomes. A larger multiplier can quickly respond to violated constraints but at the cost of increased RB consumption. Compared to these baselines, \emph{Prop-Learn} performs close to \emph{Prop-Num-MU} with a very low fraction of constraint violations.

\section{Conclusions}
In this paper, we studied the optimization of joint allocation of power and RB allocation for uplink OFDMA systems, aimed at minimizing the total number of occupied RBs subject to the QoS requirements for both LBT and SBT. We first proposed a low-complexity hierarchical algorithm to find the optimal solution in the single-user scenario, and then extended this method to address the multiuser scenario. Subsequently, we proposed a learning based method to solve the MINLP problem, where an adaptive smoothing method and a nonlinear penalty method on constraint violation  were designed. Simulation results demonstrated the superiority of the proposed method over existing methods, not only in reducing the number of consumed RBs but also in satisfying the QoS requirements with high reliability.

\begin{appendices}
\section{Derivations for Property 3} \label{proof_he}
\setcounter{equation}{0}
\renewcommand\theequation{A.\arabic{equation}}

It can be found from \eqref{posi_e} and \eqref{posi_u} that when $N^S$ is given, the values of $X^{lb}$ and $X^{ub}$ depend on $\gamma_{\min}^{L}$ and $\gamma_{\min}^{S}$. By comparing \eqref{posi_e} and \eqref{target_value}, we can obtain that
\begin{align} \label{E:XLR1}
	\begin{cases}
		X^{lb} < X^*, & \text{if}\ \gamma_{\min}^{L} > \psi,\\
		X^{lb} \geq X^*, & \text{if}\ \gamma_{\min}^{L} \leq \psi,
	\end{cases}
\end{align}
where $\psi\triangleq\frac{N}{P_{\max}+\sum_{f\in \mathcal{N}}\frac{1}{\gamma_f}}$. Similarly, by comparing \eqref{posi_u} and \eqref{target_value}, we can obtain that
\begin{align} \label{E:XLR2}
	\begin{cases}
		X^{ub} < X^*, & \text{if}\ \gamma_{\min}^{S} < \psi,\\
		X^{ub}\geq X^*, & \text{if}\ \gamma_{\min}^{S} \geq \psi.
	\end{cases}
\end{align}

Although the exact values of $\gamma_{\min}^{L}$ and $\gamma_{\min}^{S}$ are unknown since they depend on the RB allocation result $x_f^S$ as defined in \eqref{water_level_positive2a} and \eqref{water_level_positive2b}, we know that either $\gamma_{\min}^{L}$ or $\gamma_{\min}^{S}$  equals to the minimal channel gain of the $N$ RBs, denoted by $\gamma_{\min}$. We discuss the values of $\gamma_{\min}^{L}$ and $\gamma_{\min}^{S}$ in two cases.
\begin{itemize}
	\item[i)] $\gamma_{\min}^{L} = \gamma_{\min}$: In this case, the RB with the smallest channel gain is assigned to the LBT, and $\gamma_{\min}^{S} > \gamma_{\min}$ holds. Then, we can obtain from \eqref{E:XLR1} and \eqref{E:XLR2}~that
	\begin{align} \label{E:XLR3}
		\begin{cases}
			X^{lb} < X^*< X^{ub}, & \text{if}\ \gamma_{\min}^{L} = \gamma_{\min} > \psi,\\
			X^* \leq X^{lb}, & \text{if}\ \gamma_{\min}^{L} = \gamma_{\min} \leq \psi.
		\end{cases}
	\end{align}
	
	\item[ii)] $\gamma_{\min}^{S} = \gamma_{\min}$: In this case, the RB with the smallest channel gain is assigned to the SBT, and $\gamma_{\min}^{L} > \gamma_{\min}$ holds. Similarly, we can obtain that
	\begin{align} \label{E:XLR4}
		\begin{cases}
			X^{lb} < X^*< X^{ub}, & \text{if}\ \gamma_{\min}^{S} = \gamma_{\min} > \psi,\\
			X^{ub} \leq X^*, & \text{if}\ \gamma_{\min}^{S} = \gamma_{\min} \leq \psi.
		\end{cases}
	\end{align}
\end{itemize}

By combining \eqref{E:XLR3} and \eqref{E:XLR4}, we can obtain Property 3.
	
%

\section{Smoothing Parameter for Indicator Function} \label{proof_gradient}
\setcounter{equation}{0}
\renewcommand\theequation{B.\arabic{equation}}
We first examine the relationship between $|\nabla_{\tilde{g}}\tilde{\pmb{1}}{(\tilde{g},v)}|=\frac{2v e^{-v \tilde{g}}}{( 1+e^{-v \tilde{g}})^2}$ and the smoothing parameter $v$. The gradient of  $|\nabla_{\tilde{g}}\tilde{\pmb{1}}{(\tilde{g},v)}|$ with respect to $v$ is derived as
\begin{equation}\label{grad2_indi}
	\nabla_{v}\big(|\nabla_{\tilde{g}}\tilde{\pmb{1}}{(\tilde{g},v)}|\big) = \frac{2e^{v \tilde{g}}\big(e^{v \tilde{g}}+v \tilde{g}-v \tilde{g}e^{v \tilde{g}}+1\big)}{\big( 1+e^{v \tilde{g}}\big)^3}.
\end{equation}
Since $\frac{e^{v \tilde{g}}}{(1+e^{v \tilde{g}})^3}$ in \eqref{grad2_indi} is positive, the sign of  $\nabla_{v }(|\nabla_{\tilde{g}}\tilde{\pmb{1}}( \tilde{g},v)|)$ is the same as that of the term $\varphi\triangleq e^{v \tilde{g}}+v \tilde{g}-v \tilde{g}e^{v \tilde{g}}+1$.
$\varphi$ is concave with respect to $v$ since $\nabla^2_{v}\varphi=-\tilde{g}^2e^{v\tilde{g}}(v\tilde{g} + 1)\leq0$. Noting that $\varphi|_{v=0}>0$ and $\lim_{v  \to +\infty} \varphi<0$, we know that $\varphi$ is positive at first and becomes negative as $v$ increases. Thus,  $|\nabla_{\tilde{g}}\tilde{\pmb{1}}{(\tilde{g},v)}|$  first increases and then decreases with $v$.

By letting \eqref{grad2_indi} equal to zero, we know that the maximum of $|\nabla_{\tilde{g}}\tilde{\pmb{1}}{(\tilde{g},v)}|$ is achieved when $\varphi=0$.
Denoting $\zeta=v \tilde{g}$, then $\varphi=e^{\zeta}+\zeta-\zeta e^{\zeta}+1$.
The value of $\zeta$ that makes $\varphi=0$ can be found based on the following properties of $\varphi$.
First, it can be readily proved that $\varphi$ is positive when $\zeta$ is small and then becomes negative as $\zeta$ increases. Second, $\varphi$ decreases with $\zeta$ within $(1,+\infty)$, where $\varphi|_{\zeta=1}>0$ and $\lim_{\zeta  \to +\infty} \varphi<0$. These properties allow us to employ a bisection algorithm to find the value of $\zeta$ within $(1,+\infty)$ that makes $\varphi=0$ The result is $\zeta\approx1.54$. Then, we can obtain that the maximum of $|\nabla_{\tilde{g}}\tilde{\pmb{1}}{(\tilde{g},v)}|$ is $\frac{2\zeta e^{-\zeta}}{\tilde{g}(1+e^{-\zeta})^2}$, which can be achieved when $v = \frac{\zeta}{\tilde{g}}$.

Now we solve problem \eqref{P:smooth_indi}. If the maximum of $|\nabla_{\tilde{g}}\tilde{\pmb{1}}{(\tilde{g},v)}|$, i.e., $\frac{2\zeta e^{-\zeta}}{\tilde{g}(1+e^{-\zeta})^2}$, is larger than $V$, then the constraint \eqref{P:smooth_indi_c} has a solution and problem \eqref{P:smooth_indi} is feasible. Recalling that  $|\nabla_{\tilde{g}}\tilde{\pmb{1}}{(\tilde{g},v)}|$ first increases and then decreases with $v$, we know that there are two values of $v$ that can make $|\nabla_{\tilde{g}}\tilde{\pmb{1}}{(\tilde{g},v)}|=V$, which are located in $(0, \zeta/\tilde{g})$ and $ (\zeta/\tilde{g},+\infty)$, respectively. Further considering that the approximation error $|\tilde{\pmb{1}}{(\tilde{g},v)}-\pmb{1}{(\tilde{g})}|$ decreases with $v$, we know that the optimal $v$ should be located in $ (\zeta/\tilde{g},+\infty)$, which can be readily found by a bisection algorithm. If the maximum of $|\nabla_{\tilde{g}}\tilde{\pmb{1}}{(\tilde{g},v)}|$ is less than $V$,
then there does not exist a smoothing parameter $v$ satisfying constraint in \eqref{P:smooth_indi_c}  
and problem \eqref{P:smooth_indi} is infeasible. In this case, $v$ is designed to maximize the gradient $|\nabla_{\tilde{g}}\tilde{\pmb{1}}{(\tilde{g},v)}|$ and the optimal $v$ is $v=\frac{\zeta}{\tilde{g}}$.

\section{Smoothing Parameter for Piecewise\\  Maximal Function} \label{proof_gradient2}
\setcounter{equation}{0}
\renewcommand\theequation{C.\arabic{equation}}

As analyzed above \eqref{P:smooth_pm}, the desirable $\pmb{u}$ that minimizes the approximation error will make $\sum_{i\in\{L,S\}}\sum_{j=1}^{M}e^{u^i_j(p^i_{j}-p^s_{m})}\!\to\!\infty$. A large $\sum_{i\in\{L,S\}}\sum_{j=1}^{M}e^{u^i_j(p^i_{j}-p^s_{m})}$ will make the second term in \eqref{E:gradient} very small and ignorable. Thus,
the constraint \eqref{P:smooth_pm1} can be approximated for the case $p^s_{m} < \max\{\mathcal{P}^s_{m}\}$~as
\begin{equation}\label{P:smooth_pm3} \frac{1}{\sum_{i\in\{L,S\}}\sum_{j=1}^{M}e^{u^i_j(p^i_{j}-p^s_{m})}}=\bar{V}.
\end{equation}


Let $\Delta$ denote the left-hand side of \eqref{P:smooth_pm3}.
The denominator of  $\Delta$ is a sum of exponential functions $e^{u^i_j(p^i_{j}-p^s_{m})}$. Given that $u_{j}^{i} \geq 0$, $e^{u^i_j(p^i_{j}-p^s_{m})}$ is minimized by the following solution of  $\pmb{u}$ (denoted as $\pmb{u}^*$)
\begin{equation} \label{E:opt-u}
	u_{j}^{i*}=	\begin{cases}
		\frac{-\rho}{{p}_{j}^i-{p}_{m}^s}, & {p}^i_{j}<{p}^s_{m},\\
		0 , &\mathrm{else},
	\end{cases}
\end{equation}
where $\rho$ is a large constant to make $e^{-\rho}$ close to zero.
With $\pmb{u}^*$, we obtain the minimum of $e^{u^i_j(p^i_{j}-p^s_{m})}$ as 0 if ${p}^i_{j}<{p}^s_{m}$ and 1 if ${p}^i_{j}>{p}^s_{m}$. Therefore, the maximum of  $\Delta$ can be obtained as $\frac{1}{1+|\mathcal{G}|}$, where $\mathcal{G}$ denotes the set $\{{p}^i_{j}|{p}^i_{j}>{p}^s_{m},i\in\{L,S\},j=1,\ldots,M\}$, and $|\mathcal{G}|$ is the size~of~$\mathcal{G}$.

If $\frac{1}{1+|\mathcal{G}|}\geq \bar{V}$, then the optimal $\pmb{u}$ should make the constraint in \eqref{P:smooth_pm3} holding with equality. One solution is to let $e^{u^i_j(p^i_{j}-p^s_{m})}=\frac{1/\bar{V}-2M+|\mathcal{G}|}{|\mathcal{G}|}$ if ${p}^i_{j}>{p}^s_{m}$ and let $e^{u^i_j(p^i_{j}-p^s_{m})}=1$ if ${p}^i_{j}<{p}^s_{m}$. This solution can be expressed as
\begin{equation}\label{E:opt-u2}
u_{j}^i= \begin{cases}
		\frac{1}{{p}_{j}^i-{p}_{m}^s}\log {\frac{1/\bar{V}-2M+|\mathcal{G}|}{|\mathcal{G}|}},& {p}^i_{j}>{p}^s_{m},\\
		0, &\mathrm{else}.
	\end{cases}
\end{equation}
If $\frac{1}{1+|\mathcal{G}|} < \bar{V}$, then the optimal $\pmb{u}$ is $\pmb{u}^*$ given by \eqref{E:opt-u}, which maximizes $\Delta$.

By combining \eqref{E:opt-u} and \eqref{E:opt-u2}, we obtain the results in \eqref{E:opt-u0}.

\end{appendices}

\bibliographystyle{IEEEtran}
\bibliography{IEEEabrv,refer}

\end{sloppypar}
\end{document}